\documentclass[twocolumn,times]{aastex6}
\usepackage{amsmath}
\usepackage{amssymb}
\usepackage{mathtools}
\usepackage{microtype}
\usepackage{graphicx}
\usepackage[all]{hypcap}
\usepackage{needspace}

\newcommand{\acronym}[1]{\textsc{#1}}

\graphicspath{{figures/}}

 \setcitestyle{notesep={ }}

\begin{document}

\title{Ephemeris errors and the gravitational wave signal: \\ Harmonic mode coupling in pulsar timing array searches}
\author{Elinore Roebber\hyperlink{email}{$^{\star}$}}
\affiliation{School of Physics and Astronomy and Institute for Gravitational Wave Astronomy, \\
    University of Birmingham, Edgbaston, Birmingham B15 2TT, United Kingdom}
\email[\hypertarget{email}{$^{\star}$}]{e.roebber@bham.ac.uk}

\shorttitle{Ephemeris errors, GWs, and harmonic mode coupling}
\shortauthors{E.\ Roebber}

\begin{abstract}
    Any unambiguous detection of a stochastic gravitational wave background by a pulsar timing array will rest on the measurement of a characteristic angular correlation between pulsars.  The ability to measure this correlation will depend on the geometry of the array.  However, spatially correlated sources of noise, such as errors in the planetary ephemeris or clock errors, can produce false-positive correlations.  The severity of this contamination will also depend on the geometry of the array.  This paper quantifies these geometric effects with a spherical harmonic analysis of the pulsar timing residuals. At least 9 well-spaced pulsars are needed to simultaneously measure a gravitational wave background and separate it from ephemeris and clock errors.  Uniform distributions of pulsars can eliminate the contamination for arrays with large numbers of pulsars, but pulsars following the galactic distribution of known millisecond pulsars will always be affected.  We quantitatively demonstrate the need for arrays to include many pulsars and for the pulsars to be distributed as uniformly as possible.  Finally, we suggest a technique to cleanly separate the effect of ephemeris and clock errors from the gravitational wave signal. 
\end{abstract}

\section{Introduction}

Pulsar timing arrays (\acronym{pta}s) are gravitational wave (\acronym{gw}) detectors composed of radio telescopes on the earth and a collection of millisecond pulsars across the sky.  Each pulsar regularly emits radio pulses with periods of order milliseconds.  These periods are extremely stable—comparable to atomic clocks over timescales of years to decades \citep{lorimer08}.

These millisecond pulsars are used as clocks to probe nanohertz gravitational perturbations.  \acronym{Gw}s can be found by comparing the difference between the expected and actual times of arrival of pulses (`timing residuals') for multiple pulsars.

The primary signal in the \acronym{pta} band is expected to be produced by a large population of slowly-inspiralling binary systems of supermassive black holes \citep{rajagopal95}.  Such systems likely form in the aftermath of galaxy mergers \citep{begelman80}, and are thought to be numerous enough to lead to nearly-stochastic \acronym{gw} signal with a large degree of source confusion \citep{sesana08}.  

This astrophysical signal is commonly referred to as the stochastic gravitational wave background (hereafter \acronym{gwb}), and is often assumed to take the form of a statistically unpolarized and isotropically distributed Gaussian random field.   
Other possible signals, such as a background produced by inflation, may have similar statistics, but with a different spectrum \citep[e.g.][]{lasky16}.  For simplicity, this paper will assume signals with these statistical properties.

There are currently three main \acronym{pta} consortia: the European Pulsar Timing Array \citep[\acronym{epta};][]{lentati15}, the North American Nanohertz Observatory for Gravitational Waves \citep[\acronym{nanog}rav;][]{arzoumanian18}, and the Parkes Pulsar Timing Array \citep[\acronym{ppta};][]{shannon15}, all of which cooperate to form the International Pulsar Timing Array \citep[\acronym{ipta};][]{verbiest16}.  Upper limits from each \acronym{pta} are within a factor of a few of each other, and are now cutting into an astrophysically-interesting range of parameter space \citep{taylor17,arzoumanian18,chen18a}.

Looking into the future, pulsar surveys, especially those performed by the forthcoming the Square Kilometer Array, are expected to find many more millisecond pulsars \citep{keane15}.  Furthermore, new regional \acronym{pta}s in India \citep{joshi18} and China \citep{lee16} are being formed, and new telescopes will contribute to pulsar discovery and timing \citep{hobbs17}.  

\acronym{Pta}s are affected by a large number of noise sources, including intrinsic white and red noise and extrinsic and instrumental effects such as dispersion measure variation, ephemeris errors, and clock errors \citep{edwards06,cordes13}.
Although many sources of noise can be largely removed by cross-correlating timing residuals from different pulsars, there are also sources of noise which are correlated between pulsars.  Important sources of correlated noise include errors in the clock used to calibrate timing residuals and errors in the solar system ephemeris used to correct for the earth's orbit around the solar system barycenter \citep{foster90,tiburzi16}.  

Since these correlated sources of noise induce large-scale correlations between pulsars, they can mimic the effects of gravitational waves.  The ephemeris errors are of particular concern, since current measurements of the location of the solar system barycenter are not sufficiently accurate \citep{champion10,lazio18}.  Recently \cite{arzoumanian18} have shown that using recent ephemerides in a \acronym{pta} \acronym{gw} search can produce a systematic bias on the upper limit of the \acronym{gwb} amplitude.

It is well known that clock errors produce instantaneously monopolar signals, ephemeris errors produce instantaneously dipolar signals, and \acronym{gw}s produce signals which contain quadrupolar and higher order terms \citep{foster90,tiburzi16}.  However, the underlying geometry of this contamination has not yet been completely explored.  

In this paper, we will build on these well known facts to show how the contamination of the gravitational signal by the clock and ephemeris errors can be expressed in a spherical harmonic analysis of pulsar timing residuals. This framework can be used to make decisions about which pulsars should be included in \acronym{pta}s and to guide efforts to separate the signal due to \acronym{gw}s from that produced by clock or ephemeris errors.

Previous work to model and remove ephemeris errors has included simple geometric models \citep{foster90}, detailed models of the solar system \citep[e.g.][]{taylor17a}, and algebraic removal techniques based on the principles of time-delay interferometry \citep{tinto18}.  The framework proposed in this paper is related to that in \cite{tinto18}, but differs in our geometric focus and description of the origin of the contamination.  

For simplicity, we will limit the discussion to a single bin of frequency space.  The focus of this paper is on the spatial geometry, which is orthogonal to time--frequency formulations, so analysis can be straightforwardly extended to multiple frequency bins or a time-domain analysis.  \added{However, this analysis relies on simultaneous combinations of data from many pulsars, so it will not be trivial to extend it to the case of unevenly sampled data. }

The organization of this paper is as follows. \autoref{sec:ptas} presents a basic description of \acronym{pta} measurements of \acronym{gw}s. \autoref{sec:harmonic_analysis} reviews previous work by the author on using harmonic analysis to describe ideal \acronym{pta} measurements. \autoref{sec:Kllmm} shows how incomplete coverage of the sky by pulsars in the \acronym{pta} leads to coupling between different spatial modes. \autoref{sec:pta_coupling} shows how different configurations of \acronym{pta}s affect the coupling.  \autoref{sec:dealing_with_Kllmm} discusses how knowledge of the coupling can be used to remove contamination during a search for \acronym{gw}s.  \autoref{sec:conclusion} contains the conclusion.

\Needspace{7\baselineskip}
\section{Pulsar timing arrays}
\label{sec:ptas}

A gravitational wave signal can be written in the form
\begin{align}
h_{ab}(t, \vec{x}) = \int_{-\infty}^\infty \mathrm{d}f \int \mathrm{d}^2 \Omega_{\hat{n}} \, 
h_{ab}(f, \hat{n}) \, e^{2\pi i f(t + \hat{n}\cdot\vec{x}/c)},
\end{align}
where $h_{ab}$ represents a perturbation in the metric describing the curvature of spacetime, $a$ and $b$ are indices representing spatial coordinates, $\hat{n}$ gives the direction from an observer towards the \acronym{gw} source, and $\mathrm{d}^2\Omega_{\hat{n}}$ represents an infinitesimal region of sky in the direction $\hat{n}$.  The \acronym{gw}s will delay or advance pulses from a given pulsar by an amount given by \citep{romano17}:
\begin{align}
\Delta T(t, \hat{p}) = 	& \int_{-\infty}^{\infty} \mathrm{d}f \int \mathrm{d}^2 \Omega_{\hat{n}} 
					 \, \frac{1}{2} \frac{p^a p^b}{1+\hat{n}\cdot\hat{p}}  
					\,  h_{ab}(f,\hat{n}) \, e^{2 \pi i ft}\nonumber \\
		& \times \frac{1}{2\pi i f}  \left[ 1 - e^{-2\pi i f L(1+\hat{n}\cdot\hat{p})/c} \right].
\end{align}
Here, $\hat{p}$ gives the direction to the pulsar, $L$ is the distance to the pulsar, and the observer is assumed to be at the location of the solar system barycenter.  

This equation can be broken into two terms, with one subtracted from the other.  The first term, with no dependence on $L$, is known as the earth term, and represents the effect of the \acronym{gw}s on pulses arriving at the earth.  The second term, which is dependent on $L$, is known as the pulsar term, and represents the effect of the \acronym{gw}s on pulses leaving the pulsar.  If $L$ is known to a precision better than the \acronym{gw} wavelength, the pulsar term can be used to improve sky localization of individual \acronym{gw} sources \citep{zhu16}.  However, this is not the case for most millisecond pulsars.  Moreover, for a stochastic \acronym{gwb}, the pulsar term averages to a random phase and can be treated as a noise term.  For simplicity, we will neglect the pulsar term, and work with the earth term timing residuals:
\begin{equation}
r(t, \hat{p}) =  \int_{-\infty}^{\infty} \mathrm{d}f \int \mathrm{d}^2 \Omega_{\hat{n}}  \, 
			\frac{1}{2\pi i f} \, \frac{1}{2} \frac{p^a p^b}{1+\hat{n}\cdot\hat{p}} \, 
			h_{ab}(f,\hat{n}) \, e^{2 \pi i ft}
\label{eq:timing-residuals}
\end{equation}

In a single pulsar, it will be impossible to distinguish between this \acronym{gw} signal, and the timing noise of the pulsar.  Instead, a \acronym{gw} search looks for a correlated signal between different pulsars:
\begin{equation}
    \left\langle r(\hat{p}_i) r^*(\hat{p}_j) \right\rangle  = P_\textsc{gwb}(f) \, C(\vartheta_{ij}),
    \label{eq:map2pt}
\end{equation}
where $P_\textsc{gwb}(f)$ is the frequency power spectrum of the \acronym{gwb}, and $C(\vartheta_{ij})$ is the characteristic angular correlation pattern of \acronym{gw}s \citep{hellings83}:
\begin{equation}
C(\vartheta_{ij}) =\frac{1}{2} + \frac{3}{4}(1 - \cos \vartheta_{ij})  \left[\ln \left(\frac{1 - \cos\vartheta_{ij}}{2}\right)- \frac{1}{6}\right] + \frac{1}{2}\delta_{ij}
\label{eq:HD}
\end{equation}
where $\vartheta_{ij}$ represents the angle between pulsars $i$ and $j$.  This is known as the Hellings and Downs curve or the overlap reduction function.  It has been normalized so that $C(0)$ = 1, as will subsequent angular correlation functions.

The timing residuals will also be affected by an error in the terrestrial time standard, which is used to measure the time between pulses \citep{tiburzi16}.  Any error in this will affect all pulsars measured at the same time:
\begin{equation}
    r^\text{clk}(t, \hat{p}) = e^\text{clk}(t),
\end{equation}
where $e^\text{clk}(t)$ is the clock error as a function of time.  Since all pulsars are affected equally, a clock error will have a monopolar characteristic correlation pattern
\begin{equation}
    C(\vartheta_{ij}) = 1.
\end{equation}

Similarly, since \autoref{eq:timing-residuals} is calculated assuming that observations are taken at the solar system barycenter, it is necessary to translate pulse times of arrival to the solar system barycenter frame when calculating the timing residuals.  If the planetary ephemerides are not known correctly, this conversion will induce an extra term in the timing residuals \citep{tiburzi16}:
\begin{equation}
    r^\text{eph}(t, \hat{p}) = \frac{1}{c} \mathbf{e}^\text{eph}(t) \cdot \hat{p} ,
\end{equation}
where $c$ is the speed of light and $\mathbf{e}^\text{eph}(t)$ is the error in the position of the solar system barycenter with respect to time.  The characteristic correlation function for this is dipolar: 
\begin{equation}
    C(\vartheta_{ij}) = \cos \vartheta_{ij}
\end{equation}

Under ideal conditions, the characteristic correlation functions for the \acronym{gw} signal and the clock and ephemeris errors are different enough to separate out these effects.  However, this does not seem to be true in practice.  We will explore why this occurs in \autoref{sec:Kllmm} and  \autoref{sec:pta_coupling}.

\section{Harmonic space analysis}
\label{sec:harmonic_analysis}

Measuring the Hellings and Downs curve in the cross-correlation of pulsar timing residuals will be crucial to the detection of \acronym{gw}s by \acronym{pta}s. 
A previous paper by the author \citep{roebber17} developed an equivalent analysis using harmonic-space correlations of the pulsar timing residuals\footnote{
    \cite{gair14} describe a similar harmonic-space analysis for \acronym{pta}s.  However, in that work it is the \acronym{gw} strain which is decomposed into spherical harmonics, whereas we consider the decomposition of the pulsar response (the timing residuals). The \cite{gair14} framework is suited for analyzing the properties of the \acronym{gw} signal, while our framework is suited for treating other effects which cause correlations between pulsars.}.  
The results of this paper will be based on that technique, so it is summarized here.  

\begin{figure}
    \includegraphics[trim=5pt 5pt 5pt 5pt, clip=true]{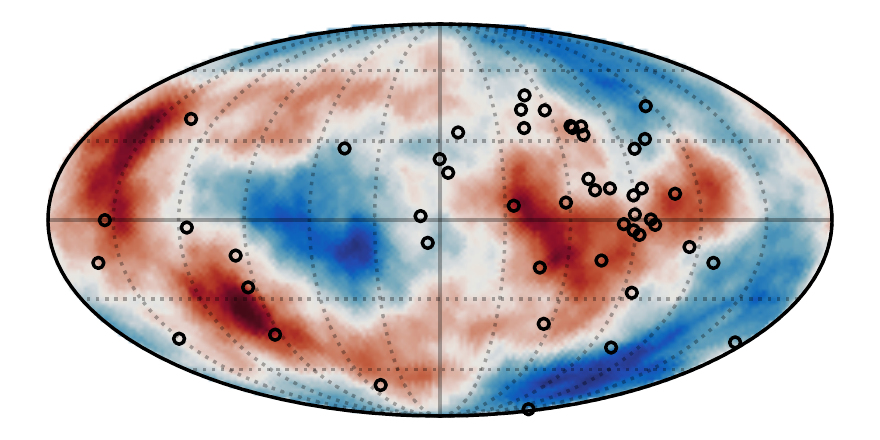}
    \includegraphics[trim=5pt 2pt 5pt 5pt, clip=true]{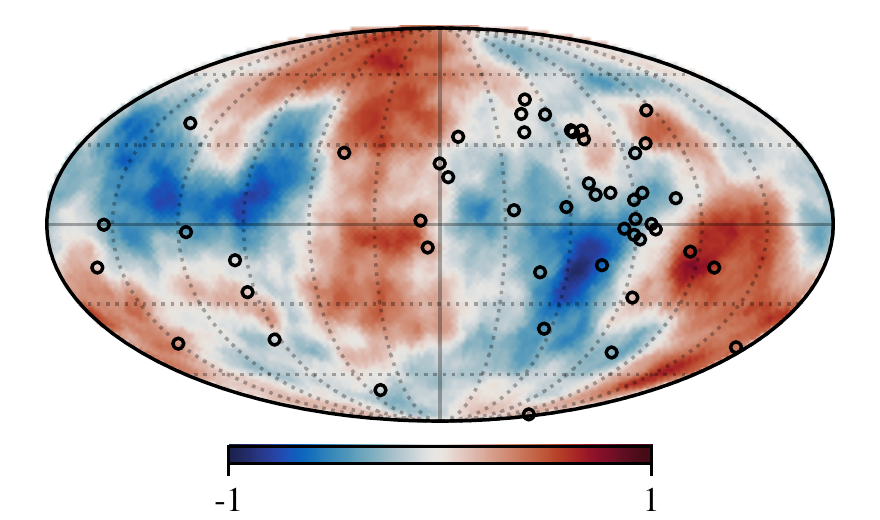}
    \label{fig:gwb-map}
    \caption{A map of the noiseless earth-term timing residuals across the sky for a single random realization of a statistically-isotropic Gaussian \acronym{gwb} in a single frequency bin. Upper and lower figures show the real and imaginary components of the complex Fourier-space timing residuals. The pulsars used in the \acronym{ipta} initial data release \citep{verbiest16} are marked with black circles. Each acquires the complex timing residual shown inside the circle.  The mostly-quadrupolar nature of the \acronym{gw} signal can be seen by eye: there are $\sim$4 large blobs of positive and negative residuals, and pulsars are correlated roughly as expected. Maps are shown in a Mollweide projection in ecliptic coordinates; units are arbitrary. 
}
\end{figure}

Consider an idealized \acronym{pta} with an infinite number of pulsars smoothly covering the entire sky.  In this case, we can make a map of the earth term timing residuals measured at each pulsar.  If the time-domain residuals are used, the map will be real and smoothly varying with time.  The frequency-domain residuals can be shown as a set of independent complex maps, with one for each frequency bin. \autoref{fig:gwb-map} shows an example of the latter case for a single frequency bin.  

A smooth, full-sky timing residual map $r(\Omega)$ can be decomposed into spherical harmonics:
\begin{align}
    r(\Omega) & = \sum_{\ell m} a_{\ell m} Y_{\ell m}(\Omega).
    \label{eq:alm-decomposition} \\
    a_{\ell m} & = \int \mathrm{d}\Omega \, r(\Omega) Y^*_{\ell m}.
    \label{eq:alm-def}
\end{align}
Here $Y_{\ell m}$ represents a spherical harmonic of degree $\ell$ and order $m$, and $a_{\ell m}$ is its amplitude.  The spherical harmonics represent a complete basis for functions defined on the sphere.

Spherical harmonics have several equivalent definitions, and are most commonly defined to be complex-valued.  However, the geometry of pulsar sky locations will be more naturally represented using the real-valued spherical harmonics.  Equations throughout this paper are appropriate for either real or complex spherical harmonics,\footnote{The exception is \autoref{eq:wignerKllmm}, which is written in terms of the real spherical harmonics.  The complex case gains some sign changes associated with the $m$ terms.} but all figures are made using real spherical harmonics.  

If the \acronym{gwb} is a Gaussian random field, each individual $a_{\ell m}$ will be random and unpredictable, but together they will be statistically described by complex Gaussian distribution. A Gaussian distribution is fully defined by two quantities: the mean and the variance (two-point correlation function).  The \acronym{gw} signal has zero mean, so the $a_{\ell m}$ distribution should as well. This leaves the two-point function as the single quantity which statistically describes the  $a_{\ell m}$ distribution.

The two-point function in harmonic space is known as the angular power spectrum and can be written \citep[e.g.][]{dodelson03} as:
\begin{equation}
    C_\ell = \frac{1}{2\ell + 1} \sum_{m=-\ell}^\ell \sum_{m'=-\ell'}^{\ell'} a_{(\ell m)} a^*_{(\ell m)'} \delta_{\ell\ell'} \delta_{mm'}.
    \label{eq:lm2pt}
\end{equation}
For a \acronym{gwb}, this becomes \citep{roebber17}:
\begin{align}
C_\ell &=
\begin{cases}
0, & \ell < 2 \\
 \dfrac{6\pi}{(\ell + 2)(\ell + 1)(\ell)(\ell - 1)}, & \ell \geq 2.
\end{cases}
\label{eq:Cl}
\end{align}

The angular power spectrum \autoref{eq:lm2pt} is mathematically equivalent to the two-point correlation function calculated in the spatial domain \citep[e.g.][]{dodelson03}:
\begin{align}
    \left\langle r(\hat{p}_i) r^*(\hat{p}_j) \right\rangle 
     = P_\textsc{gwb}(f) \sum_{\ell=0}^{\infty} \frac{2\ell + 1}{4\pi} C_\ell P_\ell (\cos\vartheta_{ij}),
    \label{eq:2pt-map-to-lm}
\end{align}
where $\vartheta_{ij}$ represents the angle between the two points $\hat{p}_i$ and $\hat{p}_j$ on the sky and $P_\ell$ is the Legendre polynomial of degree $\ell$.  $P_\textsc{gwb}(f)$ is the frequency power spectrum of the gravitational wave background.

Since \autoref{eq:2pt-map-to-lm} is equivalent to \autoref{eq:map2pt}, the angular power spectrum  contains the same information as the Hellings and Downs curve (neglecting pulsar autocorrelations):
\begin{align}
C(\vartheta_{ij}) 
& =  \sum_{\ell=0}^{\infty} \frac{2\ell + 1}{4\pi} C_\ell P_\ell (\cos\vartheta_{ij}) 
\label{eq:Cl-to-Ctheta} \\
 & = \sum_{\ell=2}^{\infty} \frac{2\ell + 1}{4\pi}  \frac{6\pi  \, P_\ell (\cos\vartheta_{ij}) }{(\ell + 2)(\ell + 1)(\ell)(\ell - 1)}  \\
 &= \frac{1}{2} + \frac{3}{4}(1 - \cos \vartheta_{ij})  \left[\ln \left(\frac{1 - \cos\vartheta_{ij}}{2}\right)- \frac{1}{6}\right].
 \label{eq:HD-no-auto}
\end{align}
This relation (\autoref{eq:Cl-to-Ctheta}) can be thought of as a spherical-harmonic version of a Fourier transform.

Both the Hellings and Downs curve (\autoref{eq:HD-no-auto}) and the angular correlation function (\autoref{eq:Cl}) represent an ensemble average over many different realizations of the \acronym{gwb}.  Any single measurement of the two-point function (in a single frequency bin) will differ from the underlying form due to sample variance \citep[see discussion in][]{roebber17}, but this can be reduced by measuring more frequency bins.

$C_\ell$ is a steeply decreasing function of $\ell$; as expected, most of the power is in the quadrupole moment ($\ell=2$), with smaller contributions from higher modes, and none in the monopole or the dipole.  This can also be seen in the example timing residual map in \autoref{fig:gwb-map}: the dominant fluctuations are those covering about $ 1/4 $ of the sky, smaller-scale fluctuations are much less important, and fluctuations on scales smaller than $ \sim 10^\circ $ are practically absent.  The higher-than-quadrupole terms are due to the factor of $(1 + \hat{n}\cdot\hat{p})^{-1}$ in \autoref{eq:timing-residuals}, and are unrelated to any non-quadrupolar components of the strain.  

The quadrupole term of $C_\ell$ contains nearly all of the signal-to noise of the total \acronym{gwb}, so a successful detection of the \acronym{gwb} will require the detection of $C_2$.  Most contributions to the noise are uncorrelated between pulsars, and should affect all terms of $C_\ell$ equally.  

The spatially correlated noise terms discussed in \autoref{sec:ptas}, however, have their own characteristic angular power spectra.  In both cases, these are very simple: clock errors produce a monopole ($\ell=0$) signal, and ephemeris errors produce a dipole ($\ell=1$) signal.  Harmonic analysis of the timing residuals naturally separates these spatially correlated noise signals from the \acronym{gw} signal ($\ell\geq2$).

Given this natural separation, why are we concerned that measurements of the \acronym{gw} signal might be contaminated by ephemeris or clock errors?  So far, we have assumed that our array of pulsars covers the sky sufficiently completely that we may accurately measure spherical harmonics up to $\ell = 2$.  However, a real \acronym{pta} will have a limited number of pulsars distributed non-uniformly across the sky, and as a result, measurements of spherical harmonics will suffer from convolution with the geometry of the pulsar array.  This effect will be the subject of the next section.

\section{The mode coupling matrix}
\label{sec:Kllmm}

We previously assumed an ideal \acronym{pta} which can measure timing residuals at every point on the sky.  However, a real \acronym{pta} is limited to only tens of points on the sky: the locations of each pulsar in the array.  This section generalizes the results from the previous section to include the effects of a \acronym{pta} which instead sparsely samples the underlying timing residual map.

In an ideal \acronym{pta}, the measured timing residuals form a smooth field, which can be straightforwardly transformed into spherical harmonics, as in \autoref{eq:alm-decomposition}.  This is possible because the spherical harmonics form an orthonormal basis for fields on a sphere.  However, for a finite number of pulsars, the spherical harmonics sampled at the locations of the pulsars will no longer form an orthogonal basis function for this space.  

\begin{figure}
    \includegraphics[trim=5pt 2pt 5pt 5pt, clip=true]{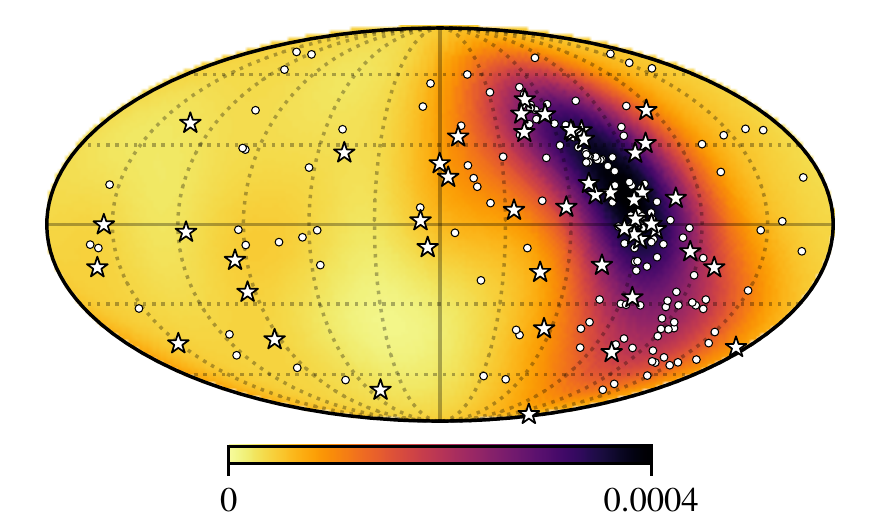}
    \caption{Sky locations of known millisecond pulsars from the \acronym{atnf} pulsar catalogue\textsuperscript{\ref{fnote}} \citep{manchester05}, in ecliptic coordinates on a Mollweide projection. Pulsars used in the first \acronym{ipta} data release \citep{verbiest16} are indicated with stars; other millisecond pulsars are marked with circles. The smoothed distribution is used throughout this paper to draw random realizations of `galaxy-distributed' pulsars.}
    \label{fig:msps}
\end{figure}

Attempting to measure the amplitudes of the standard spherical harmonics will have mixed results: some harmonics will be poorly sampled and power will bleed between different modes.  This is particularly a concern if the pulsars are not evenly distributed across the sky, as is the case for the population of known millisecond pulsars (see \autoref{fig:msps}).

The degree to which power from one mode can bleed into another can be represented by a coupling matrix  \citep{peebles73,gorski94a,wandelt01,mortlock02,hivon02,efstathiou04}, which shows how much each spherical harmonic overlaps with all other harmonics:
\begin{align}
\label{eq:coupling_matrix_int}
K_{(\ell m)(\ell m)'} &= \int \! \mathrm{d}\Omega \, Y_{\ell m}(\Omega) \, W(\Omega) \,Y^*_{(\ell m)'}(\Omega) \\
&= \frac{4\pi}{\sum_i w_i}\sum_i w_i Y_{\ell m}(\hat{p}_i) Y^*_{(\ell m)'}(\hat{p}_i).
\label{eq:coupling_matrix_sum}
\end{align}
The window function $W(\Omega)$ is used to define the part of the sky measured by pulsars:
\begin{equation}
W(\Omega) = \sum_i w_i \, \delta(\Omega - \hat{p}_i),
\label{eq:window}
\end{equation}
where the pulsar sky locations are given by $\hat{p}_i$ and the weights for each pulsar are given by:
\begin{align}
    w_i&\propto
    \begin{cases}
     \sigma_i^{-2}, & \sigma_i^{2}\text{ is the noise variance of pulsar $i$.} \\
     1, & \text{for an array of equal pulsars.}
    \end{cases}
\end{align}
We will discuss the first case (pulsars have different levels of intrinsic noise) in \autoref{sec:psr-quality}, and otherwise assume the second case (all pulsars are equal).

\footnotetext{\url{http://www.atnf.csiro.au/research/pulsar/psrcat/} (accessed May 5, 2018) \label{fnote}}

If complex spherical harmonics are used to calculate $K_{(\ell m)(\ell m)'}$, the coupling matrix will be Hermitian.   For real spherical harmonics, it will be real and symmetric.

When the integral in \autoref{eq:coupling_matrix_int} is taken over the entire sphere with $W(\Omega)=1$ everywhere, the coupling matrix becomes  $K_{(\ell m)(\ell m)'} = \delta_{\ell\ell'} \delta_{mm'}$, and the spherical harmonics are orthonormal once more. 

The properties of the coupling matrix determine whether or not we will be able to algebraically disentangle the different modes.  When the different harmonics are orthogonal, the coupling matrix is diagonal, and the problem is trivial.  In the opposite limit, where all harmonics are maximally entangled, the coupling matrix is singular, and the problem is impossible.  In general, if the coupling matrix is well-conditioned (invertible), it should be possible to disentangle the  \acronym{gw} signal from dipolar ephemeris and monopolar clock errors.  

Since the \acronym{pta} response function $\propto(1 + \hat{n}\cdot\hat{p})^{-1}$, the spherical harmonic transform of the  timing residuals will contain \acronym{gw} signal at all $\ell \geq 2$, even for a strain signal which is strictly quadrupolar.  However, since the majority of the signal is contained in the quadrupole, we will neglect these higher order terms.

Using spherical harmonic product identities, \autoref{eq:coupling_matrix_int} can also be written \citep[e.g.][]{hivon02}:
\begin{align}
K_{(\ell m)(\ell m)'} = &  \frac{4\pi}{\sum_i w_i} 
\sum_{L,M}\sum_i  
\sqrt{\frac{(2\ell + 1)(2\ell'+1)(2L + 1)}{4\pi}} \nonumber \\
& \times \left(\begin{array}{ccc} 
\ell & \ell' & L  \\ 
 m & m' & M \\ 
\end{array}\right)
\left(\begin{array}{ccc} 
\ell & \ell' & L \\ 
0 & 0 & 0 \\ 
\end{array}\right)
 w_i Y_{LM} (\hat{p}_i),
\label{eq:wignerKllmm}
\end{align}
where terms of the form
\begin{equation*}
\left(\begin{array}{ccc} 
\ell_1 & \ell_2 & \ell_3  \\ 
m_1 & m_2 & m_3 \\ 
\end{array}\right)
\end{equation*}
are known as Wigner 3j symbols, and are closely related to the Clebsch-Gordan coefficients.  They can generally not be written in closed form, but obey the following symmetry rules \citep[e.g.][]{hivon02}:
\begin{align}
& m_1 + m_2 + m_3 = 0 & &\\
& |\ell_1 - \ell_2| \leq \ell_3 \leq \ell_1 + \ell_2 & & \text{(triangle inequality)} \label{eq:wig-triangle} \\
& \ell_1 + \ell_2 + \ell_3 \text{ is even} & & \text{if } m_1 = m_2 = m_3 = 0 \label{eq:wig-even}
\end{align}
This form of the coupling matrix is useful, since it is explicitly written in terms of the pulsar sky locations $\sum_i w_i Y_{LM}(\hat{p}_i)$. 
This will allow us to gain insight into how the latter affects the former (see \autoref{sec:sky-dist}).

\section{PTA configuration and mode coupling}
\label{sec:pta_coupling}
 
$K_{(\ell m)(\ell m)'}$ is entirely determined by the properties of the pulsar array.  Important properties include the number of pulsars in the array, the spatial distribution of the pulsars, and the quality of each pulsar.

\begin{figure}
    \includegraphics[trim=5pt 5pt 5pt 5pt, clip=true]{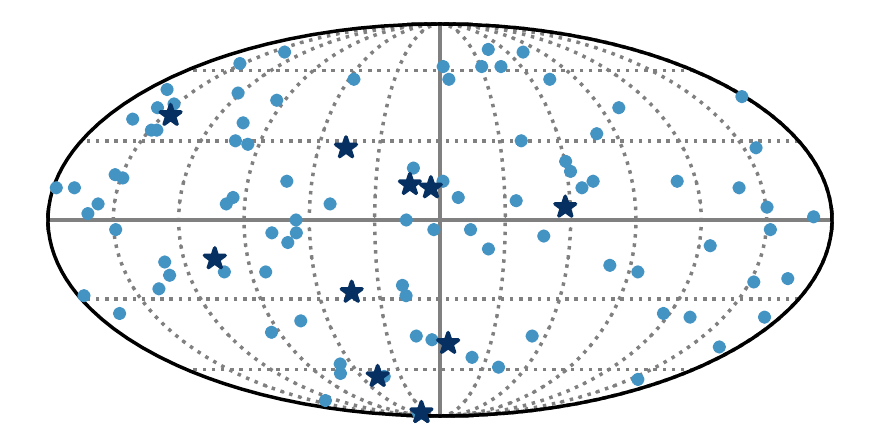}
    \includegraphics[trim=5pt 5pt 5pt 5pt, clip=true]{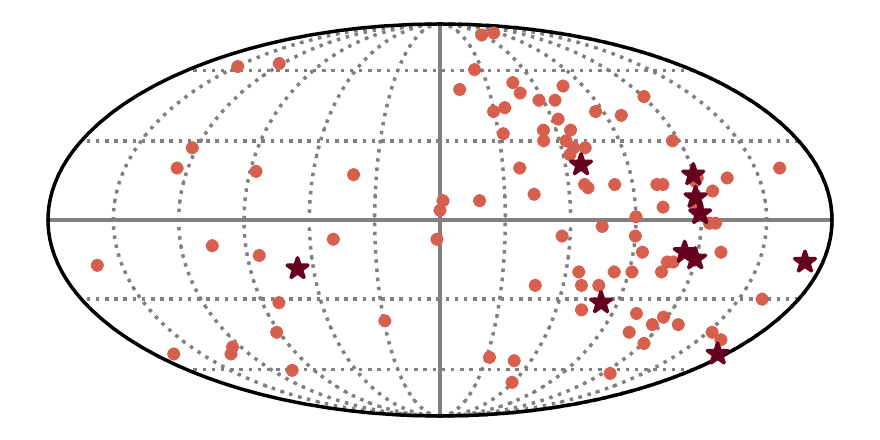}
    \label{fig:psr-distributions}
    \caption{Toy model \acronym{pta}s with 100 pulsars, shown in a Mollweide projection.  In the top example, pulsars are drawn from a uniform sky distribution.  In the bottom example, pulsars are drawn from a distribution based on the known distribution of millisecond pulsars (as in \autoref{fig:msps}).  In both cases, stars represent the subset of 10 pulsars used for both the 10-pulsar arrays and also the 10 `good' pulsar arrays.}
\end{figure}

The effects of varying these properties will be illustrated by a set of 6 toy model \acronym{pta}s.  The models include arrays with 10 pulsars, with 100 pulsars, and with 10 good pulsars (with weight 1) and 90 poor pulsars (with weight 0.05).   The pulsars in each case are generated by random sampling from two spatial distributions: `uniform' and `galaxy'.

The  `uniform' distribution represents the case where every direction is equally likely to contain a pulsar.  The `galaxy' distribution is based on the known population of millisecond pulsars, as shown in \autoref{fig:msps}.  The distribution was created by selecting millisecond pulsars from the \acronym{atnf} pulsar catalogue \citep{manchester05}, removing multiples associated with globular clusters to avoid biasing the distribution towards those sky locations, and smoothing with a Gaussian beam with full width at half maximum of $45^\circ$.  

\begin{figure}
    \includegraphics[trim=5pt 5pt 5pt 5pt, clip=true]{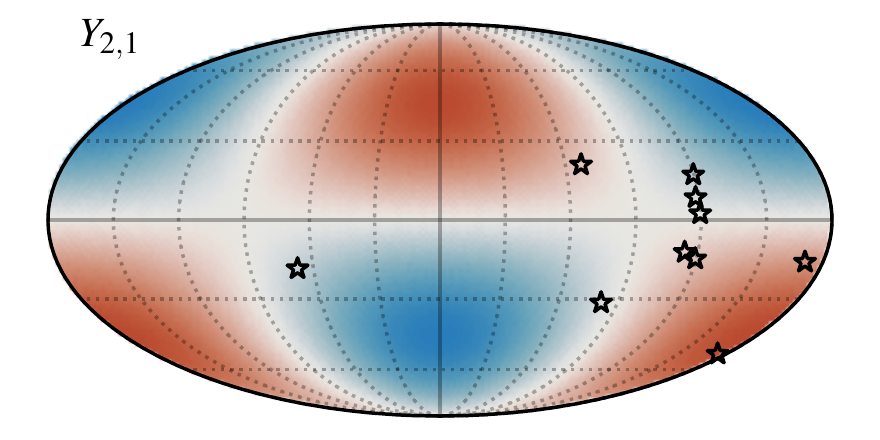}
    \includegraphics[trim=5pt 2pt 5pt 5pt, clip=true]{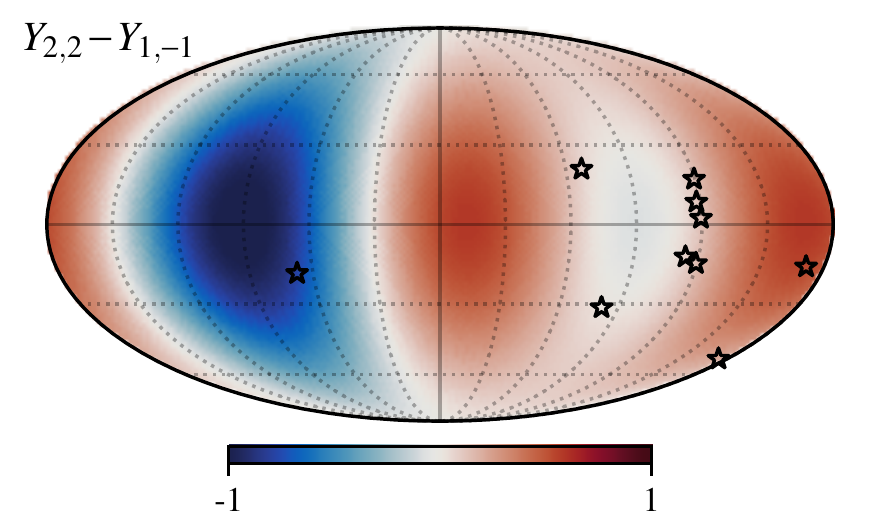}
    \caption{Examples of difficult harmonics for the 10-pulsar `galaxy' array.  The top panel shows the real $Y_{2,1}$ harmonic.  Since most pulsars lie close to one of its zeros, its amplitude will not be well-measured by the array, potentially leading to an underestimate of the \acronym{gwb} amplitude. The bottom panel shows the difference between the real $Y_{2,2}$ and $Y_{1,-1}$ harmonics.  Pulsars which are near a zero of this quantity have trouble distinguishing between the two harmonics, leading to a coupling between these modes. This potentially allows an ephemeris error to be mismeasured as a \acronym{gw} signal.}
    \label{fig:badYlm}
\end{figure}

The distribution of pulsars for the `uniform' and `galaxy' cases are shown in \autoref{fig:psr-distributions}.  The 10-pulsar subsets used for the `10 equal pulsars' and the `10 good pulsars' are marked with dark stars. Examples of harmonics which are not well measured or are coupled to other harmonics are shown in \autoref{fig:badYlm}. The coupling matrix for each model \acronym{pta} is shown in \autoref{fig:Kllmm}, and their eigenvalues in \autoref{fig:svd}.

\subsection{Pulsar number}

The first important consideration is the number of pulsars in the array, or the number of points at which the signal is sampled.   To successfully disentangle \acronym{gw} signal and spatially correlated clock and ephemeris errors, it will be necessary to measure the $\ell=0$ (monopole), $\ell=1$ (dipole), and $\ell=2$ (quadrupole) moments of the \acronym{pta} response.  With too few pulsars, some modes of interest will be poorly sampled, and $K_{(\ell m)(\ell m)'}$ will not be invertible.  This problem is closely related to the Nyquist sampling theorem.

A lower limit on the number of pulsars required can be calculated using a mode-counting argument.  There are $2\ell + 1$ harmonic modes $m$ for every multipole $\ell$.  For a frequency-domain signal, the timing residuals are complex, and all modes are independent. If we assume that the \acronym{gw} signal is approximately band-limited with $\ell \leq \ell_\text{max}$, we can count the number of modes needed:
\begin{equation}
   \sum_{\ell=0}^{\ell_\text{max}}  2\ell + 1 = (\ell_\text{max} + 1)^2.
\end{equation}

Measuring the leading-order term of the \acronym{gw} signal (the quadrupole) and disentangling it from the potentially nonzero monopole and dipole terms would therefore require a minimum of 9 pulsars.  Measuring up to the octopole term of the \acronym{pta} response would require a minimum of 16 pulsars.  

\begin{figure}
    \includegraphics{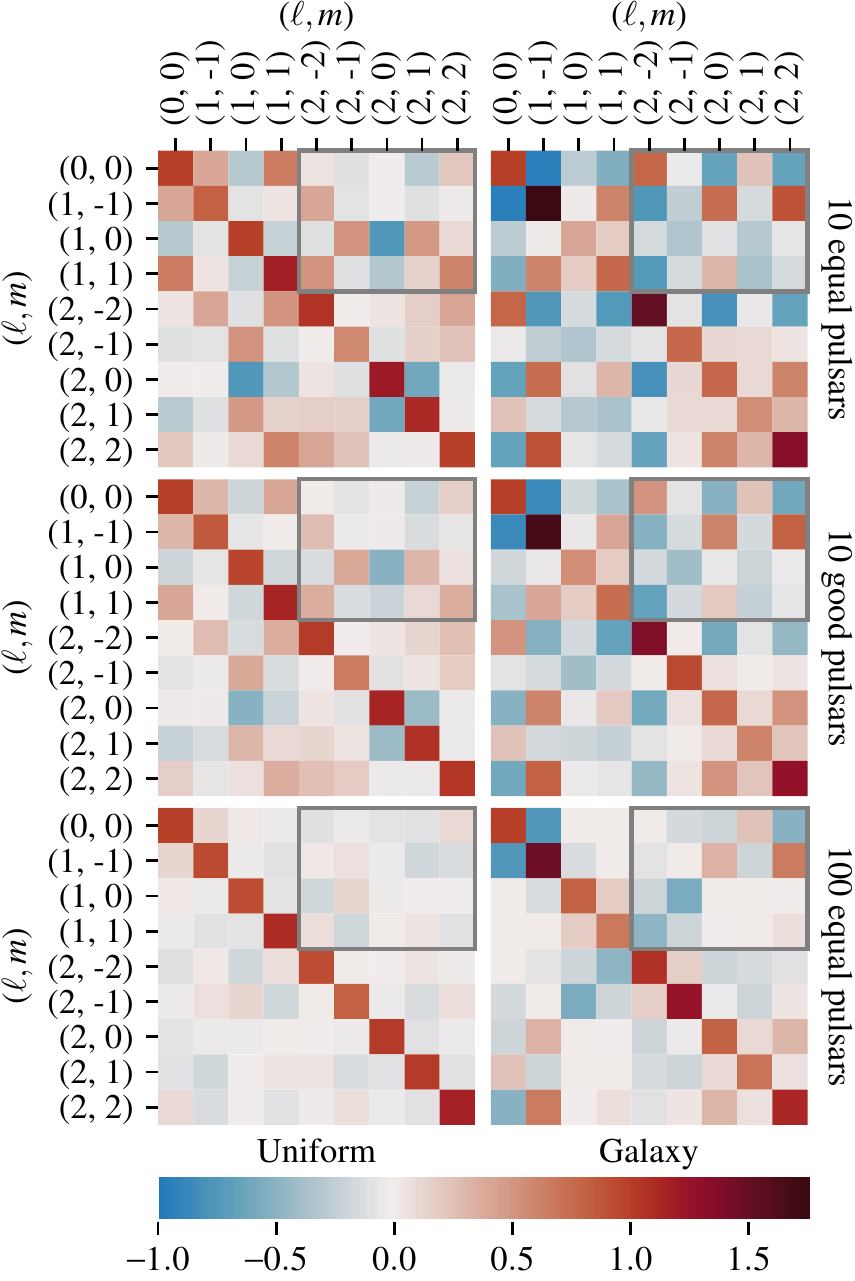}
    \label{fig:Kllmm}
    \caption{Coupling matrices $K_{(\ell m)(\ell m)'}$ for 6 different toy \acronym{pta}s with the sky distributions given in \autoref{fig:psr-distributions}. The left column shows  uniformly-sampled \acronym{pta}s, and the right shows galaxy-sampled \acronym{pta}s. The top row shows arrays with 10 pulsars.  The middle row shows arrays where the first 10 pulsars are `good' (weight~$=1$) and the other 90 are `bad' (weight~=~0.05). The bottom row shows the coupling matrix for the full equally-weighted 100-pulsar array.  The grey boxes highlight terms which represent coupling between the monopole or dipole modes and the quadrupole modes. Any non-zero terms within a box can lead to clock or ephemeris errors being mistaken for a \acronym{gw} signal.}
\end{figure}

However, larger arrays may perform significantly better.  As can be seen in \autoref{fig:Kllmm}, the coupling matrices for the 10-pulsar arrays have large off-diagonal elements, indicating strong coupling, and occasional weak diagonal elements where modes are not well measured (see also \autoref{fig:badYlm}).  These aspects improve for arrays with more pulsars.  

\begin{figure}
    \includegraphics{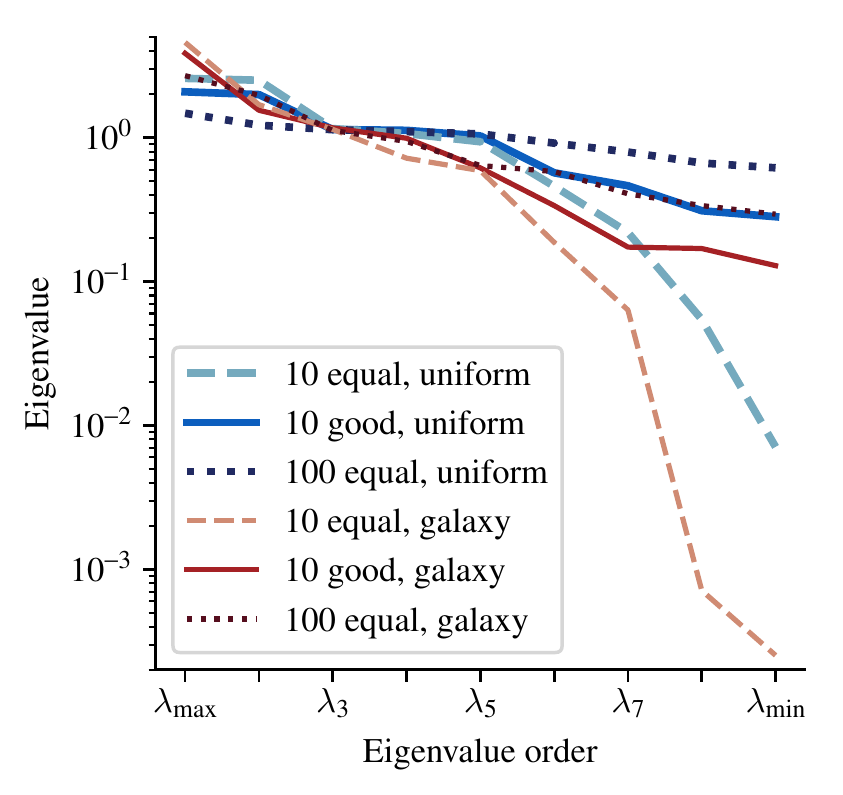}
    \caption{Sorted eigenvalues of the coupling matrices from \autoref{fig:Kllmm}.  Large arrays (dotted lines) perform well, small arrays (dashed lines) perform poorly.  Coupling matrices for arrays with 10 `good' pulsars and 90 `bad' pulsars (solid lines) are nearly as good as the equally-weighted case, despite the appearance of \autoref{fig:Kllmm}.  This is because their largest eigenvalues follow those of the 10-pulsar arrays, but the smallest are raised. Uniformly-drawn arrays (blue, thick) give better results than the arrays drawn from a galactic distribution (red, thin).}
    \label{fig:svd}
\end{figure}

To determine whether the coupling can be removed, we want to know if the coupling matrix is well-conditioned.  This can be evaluated by looking at its eigenvalue decomposition, as in \autoref{fig:svd}.

A matrix with at least one zero-valued eigenvalue is singular.  
If the eigenvalues are nonzero, but very small, it may be possible to numerically invert the matrix, but using the inverse will magnify errors in the data. In this scenario, the matrix is ill-conditioned. A useful figure of merit is the condition number: the ratio between the largest and smallest eigenvalue.  When the condition number is close to 1, the matrix is well-conditioned; when it is large, the matrix is ill-conditioned; and when it is infinite, the matrix is singular.  \added{In this paper, we will not address the issue of choosing where to draw the line between well- and ill-conditioned matrices.  The range of acceptable values will likely depend on the details of the analysis chosen, and should be assessed using simulations with a more thorough treatment of the various sources of noise than are considered here.}  

Although the mode-counting argument above suggested that arrays with $\geq9$ pulsars should have well-conditioned coupling matrices, the 10-pulsar examples seen in \autoref{fig:svd} are marginal, particularly in the case of the galaxy-distributed \acronym{pta}.  A coupling matrix for a \acronym{pta} with a small number of pulsars can have a wide range of condition numbers, depending on the locations of the different pulsars.  

\begin{figure}
    \includegraphics[]{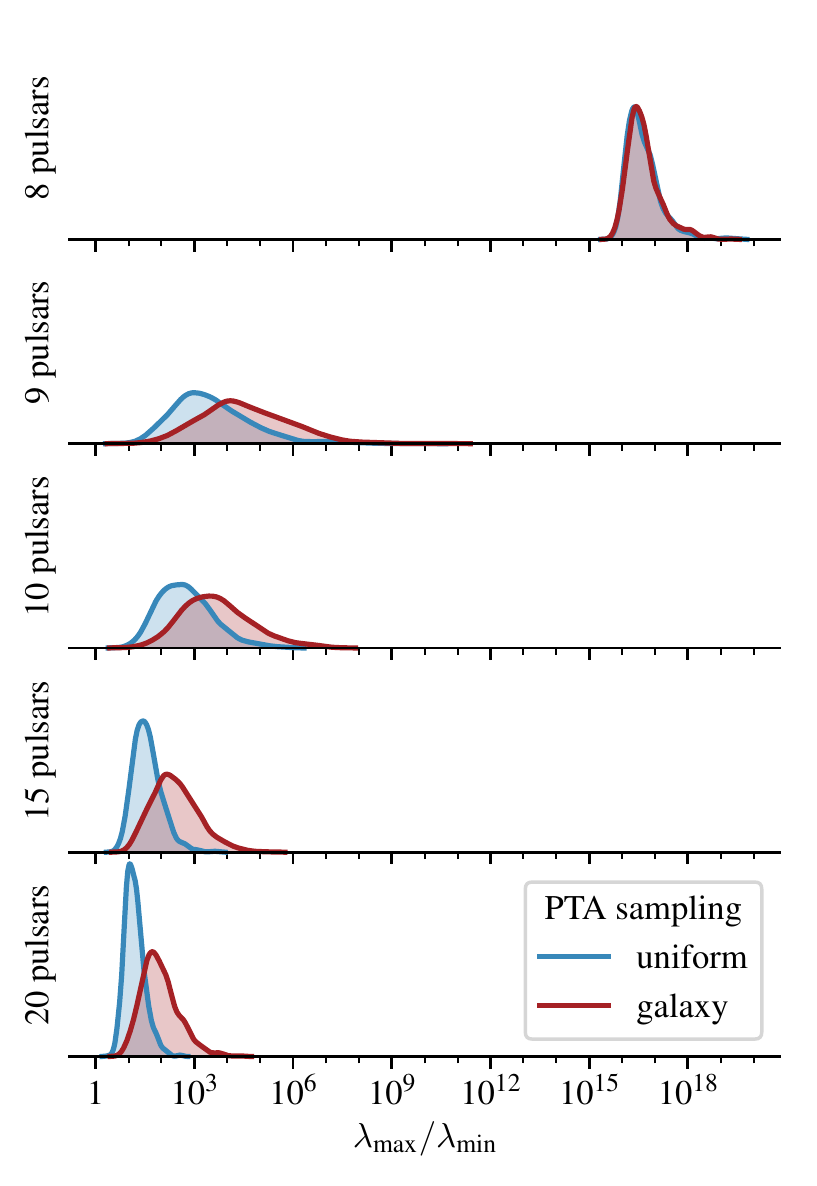}
    \caption{Distributions (using a kernel density estimator) of the ratio between largest and smallest eigenvalue of the coupling matrix for 1000 realizations each of different random \acronym{pta} configurations.  When this ratio is close to 1, the coupling matrix is well-conditioned, and can be safely inverted.  When this ratio is large (but not infinite), the inverse coupling matrix can be computed, but using it will amplify errors.  An array of 9 pulsars is the smallest array which can have a well-conditioned coupling matrix (for $\ell_\text{max} = 2$), but a larger array is more likely to.}
    \label{fig:cond}
\end{figure}

\autoref{fig:cond} shows the condition number for 1000 different random realizations of arrays with a variable number of pulsars.  From the top plot, it is clear that the coupling matrix of any array with $\leq$ 8 pulsars is numerically singular, while for larger arrays, the coupling matrix is not.  However, 9- or 10-pulsar arrays show a spread of many orders of magnitude in condition number.  Some distributions will be well-conditioned, and some will certainly not.  As the number of pulsars increases, the mean and spread decrease, and an array with $\gtrsim$~20 pulsars, will be more likely to have a well-conditioned coupling matrix.
 
This result agrees with previous work \citep{siemens13,taylor16a} arguing that large arrays are needed to detect a stochastic \acronym{gwb}, albeit for different reasons.  However, other researchers \citep{foster90,cornish16,tinto18} have suggested that a 5-pulsar array could be sufficient, which contradicts the 9-pulsar minimum found in this paper. 

In the case of \cite{foster90} and \cite{tinto18} this is due to a difference in the counting arguments used.  Their 5 pulsars can be understood to include 1 pulsar to measure and remove the monopole clock error, 3 pulsars to measure and remove the dipole ephemeris error, and a final pulsar to measure the \acronym{gwb}. However, if the \acronym{gwb} is assumed to have the form of a Gaussian-random quadrupole, it will have 5 independent components, bringing the minimum to 9 pulsars.

On the other hand, \cite{cornish16} show that small 5-pulsar arrays are adequate for detecting a sufficiently isotropic \acronym{gwb}, in the absence of correlated noise terms.  But this does not contradict our results, since they also find that small arrays perform less well for \acronym{gw} signals dominated by small numbers of sources, or in the presence of correlated noise.

The number of pulsars needed for the coupling matrix to be well-conditioned is a useful lower limit on the necessary size of a \acronym{pta}. However, as can be seen in \autoref{fig:Kllmm}, large arrays have less intrinsic coupling between the quadrupole and monopole/dipole modes than small arrays.  

This leads to another important threshold: the number of pulsars needed for the coupling between the \acronym{gw} signal and clock/ephemeris error to become unimportant.   It happens that the distribution of pulsars across the sky has a significant impact on this threshold, so it will be discussed in the next section.

\Needspace{7\baselineskip}
\subsection{Pulsar sky distribution}
\label{sec:sky-dist}

The second important aspect is the sky distribution of the pulsars.  Millisecond pulsars are not located uniformly across the sky (see \autoref{fig:msps}).  As a result, the distribution of pulsar separations will not be uniform in $\cos \alpha$, but will be biased towards smaller separations \citep[see figures in][]{verbiest16,arzoumanian18}.  Furthermore, they are predominantly located in an area which surrounds the center of the galaxy, and is also nearly centered on the ecliptic. 

The effect of the pulsar spatial distribution on the form of the coupling matrix is perhaps most clear in \autoref{eq:wignerKllmm}. In this equation, terms of the form $w_iY_{LM} (\hat{p}_i)$ contain information about the pulsars in the array. The Wigner symmetry rules (\autoref{eq:wig-triangle} and \autoref{eq:wig-even}) determine which $Y_{LM}$ terms are important for coupling.  In particular, we are interested in coupling between $\ell=2$ and $\ell=1$ modes and between $\ell=2$ and $\ell=0$ modes.  In the first case, harmonics with $L=1,3$ will contribute, and in the second case, only harmonics with $L=2$ will matter.  

This means that any pulsar timing array distribution which has significant dipole, quadrupole, or octopole components will lead to coupling between the quadrupole \acronym{gw} signal and dipole (monopole) ephemeris (clock) error.  In particular, the distribution of known millisecond pulsars correlates with the galaxy; from \autoref{fig:msps}, it is clear that distributions of pulsars will naturally tend to cluster around the center of the galaxy, and will therefore have non-negligible dipole, quadrupole, and octopole components.   As a result, \acronym{pta}s naively chosen from a collection of the best pulsars will be predisposed towards strong coupling between the quadrupolar signal and the undesired dipole and monopole sources. 
  
\begin{figure}
     \hspace{-10pt}
     \includegraphics[]{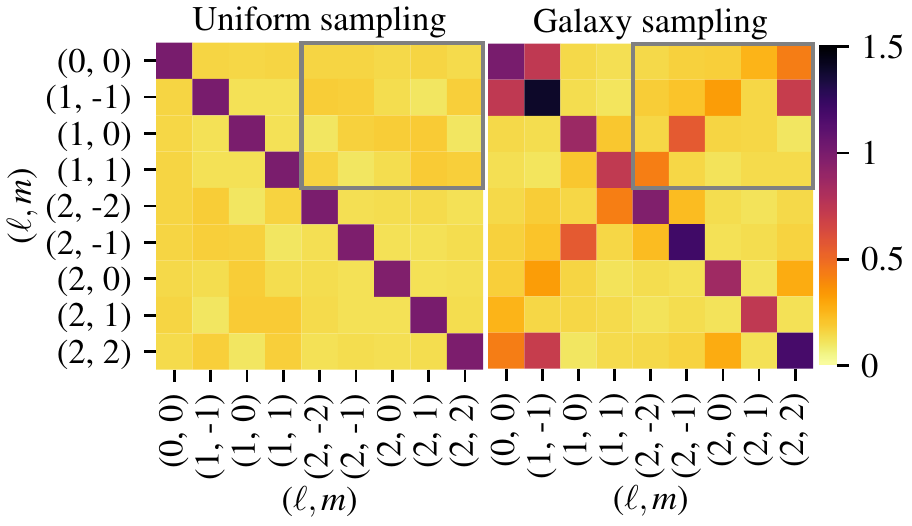}
     \caption{Median of the absolute value of the coupling matrix for 1000 realizations of 20-pulsar arrays sampled from galaxy and uniform distributions of pulsars. The grey rectangle highlights problematic coupling terms, as in \autoref{fig:Kllmm}.  Uniformly-sampled \acronym{pta}s have very little common structure in their coupling matrices, and as the number of pulsars increases, they tend towards the identity matrix.  By contrast, galaxy-sampled \acronym{pta}s show clear common structure, with significant coupling (on average) between every dipole mode and all but one quadrupole modes, as well as coupling between the monopole mode and quadrupole modes.  Although shown here for the 20-pulsar case, this structure persists for arbitrarily large numbers of galaxy-sampled pulsars.}
     \label{fig:abs_coupling_heatmap}
\end{figure}

 \autoref{fig:abs_coupling_heatmap}, which shows the underlying structure of the amplitudes of the coupling matrices, confirms this.  The uniformly-distributed case has no underlying structure, while the galaxy-distributed case has an underlying tendency for certain modes to be coupled.  
  
\begin{figure}
     \includegraphics{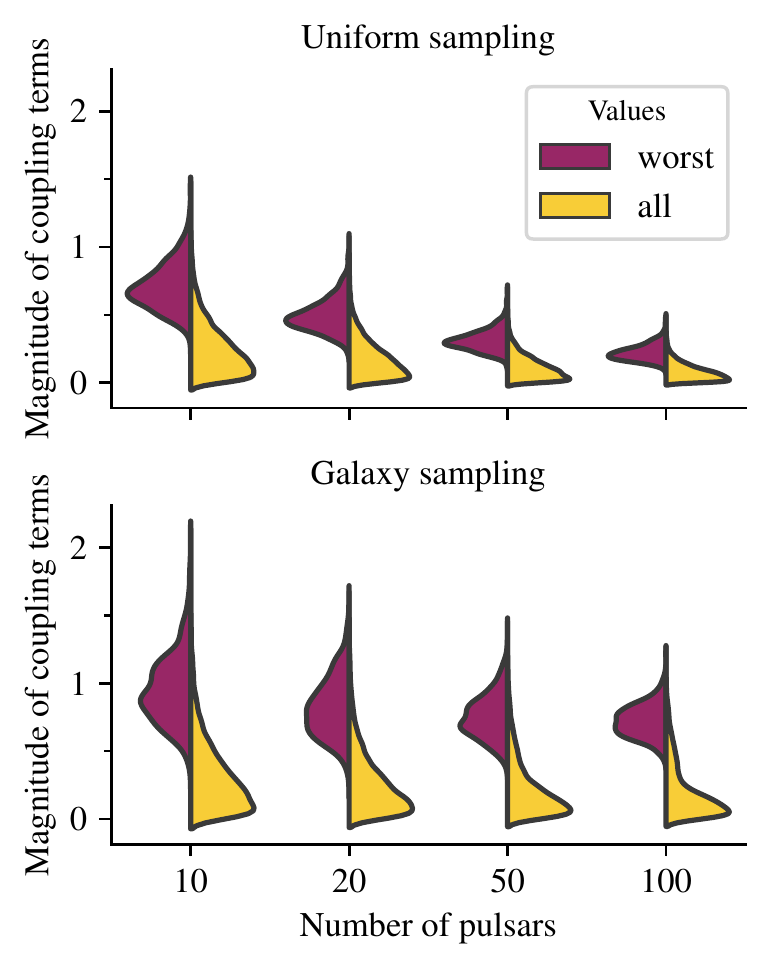}
     \caption{Violin plot of the distribution of absolute values of the monopole--quadrupole and dipole--quadrupole coupling terms (shown in boxes in \autoref{fig:Kllmm} and \autoref{fig:abs_coupling_heatmap}). Each violin represents 1000 realizations of arrays drawn from the `galaxy' or `uniform' distributions with 10, 20, 50, or 100 pulsars. Yellow curves show the full distribution of amplitudes for all 20 coupling terms in all realizations, while purple curves pick out the worst (largest absolute value) coupling term in every realization.  Galaxy-sampled \acronym{pta}s almost always have at least one coupling term $\gtrsim 0.5$, no matter how many pulsars are used, while uniformly-sampled \acronym{pta}s with $\gtrsim50$ pulsars almost never do. }
     \label{fig:abs_coupling_violin}
\end{figure}

 While \autoref{fig:abs_coupling_heatmap} shows the median coupling matrix for many different random realizations, \autoref{fig:abs_coupling_violin} shows how the amplitudes of the problematic terms of the coupling matrix are distributed across many random realizations.  This figure shows that while larger arrays have better coupling matrices, the worst coupling term for galaxy-distributed arrays remains large no matter how many pulsars are included.  This worst coupling term has a median value $\sim 0.7$, but can be much higher.  From \autoref{fig:abs_coupling_heatmap}, we can see that the strongest coupling term is typically between the $(\ell, m) = (2,2)$ mode and the $(1,-1)$ mode.
 
For galaxy-distributed \acronym{pta}s with many pulsars, most potentially-problematic terms in the coupling matrix are small, but none of the quadrupole modes is free from coupling.
The $(2,1)$ mode shows the least coupling overall ($\sim 0.2$ to the monopole), but it is also the most poorly sampled of the quadrupole modes.  These effects can also be seen in the bottom right coupling matrix in \autoref{fig:Kllmm}.  With 100 pulsars, this coupling matrix is well-conditioned, but clearly shows the same structure as \autoref{fig:abs_coupling_heatmap}. 
 
\acronym{Pta}s with uniformly distributed pulsars show similarly large couplings when the number of pulsars is small.  But as the number of pulsars in the array increases, the overall coupling between quadrupole and monopole/dipole modes decreases, and the matrix becomes increasingly diagonal.  With 50 pulsars in the array, the mode with the worst coupling is always $\lesssim 0.5$, and typical modes are $\sim 0.1$ (see \autoref{fig:abs_coupling_violin}).  
 
Futuristic \acronym{pta}s with $\gtrsim 50$ high-quality pulsars may be able to eliminate contamination by solar system ephemeris or clock errors, but only if the pulsars are distributed nearly uniformly across the sky.  Current \acronym{pta}s have been constructed to be more isotropic than our naive galaxy-distributed model, but they are not isotropic enough to have low levels of contamination.

In addition to the large-scale distribution of pulsars (dipole--octopole moments), the small-scale distribution of pulsars can also have an effect.  Tightly clustered pulsars (separations  $\ll 180^\circ/\ell_\text{max}$) will make highly correlated measurements of the large-scale modes, and so cannot really be considered to be independent.  This means that adding an additional pulsar close to one already present will not improve the coupling matrix as well as adding a widely-spaced pulsar would.  

As an example, the galaxy-distributed 10-pulsar model (shown in the bottom plot of  \autoref{fig:psr-distributions}) has two sets of pulsars which are very close to one another.  It might be expected to act somewhat like a 7-pulsar array.  This is reflected in the eigenvalues of its associated coupling matrix (\autoref{fig:svd}): there is a drop of two orders of magnitude between the seventh- and the eighth-largest eigenvalues.

However, if there are enough widely-separated pulsars in the array, tightly-clustered pulsars may not be a problem.  In particular, poor-quality pulsars with large amounts of white or red noise might not be useful on their own.  But if there are several of them all sampling the same underlying \acronym{gw} (and ephemeris error) signal, they could act similarly to a single higher-quality pulsar.  This effect could be employed in order to build a more uniform array of pulsars, with multiple poor-quality pulsars used to fill large gaps between the better ones.

In summary, a \acronym{pta} must have $\geq 9$ pulsars in order to be able to separate quadrupolar \acronym{gw} signal from dipolar ephemeris error and monopolar clock error, but there are also requirements on the distribution of those pulsars.  Ideally, the pulsars should be arranged uniformly across the sky, with separations $\sim 180^\circ/\ell_\text{max}$.  Non-uniformly distributed pulsars will increase the coupling.  An array with more closely-spaced pulsars than optimal will need to have more pulsars in order to properly sample the relevant modes.

It is worth noting that this spacing recommendation is different from the spacing that maximizes the detectability of the Hellings and Downs curve.  The most important difference is that pulsars with small separations are not useful for improving the coupling matrix.  This is because correlations between tightly coupled pulsars may be useful for detecting a correlated signal, but they will be useless for determining whether this correlation is due to a \acronym{gw} signal, an error in the planetary ephemeris, or a clock error.   

\subsection{Pulsar quality}
\label{sec:psr-quality}

The final important effect is the quality of different pulsars.  This can be treated using the weights defined in \autoref{eq:window}, so that `good' pulsars have a large weight and `bad' pulsars a small one. 

The example for this scenario has the full set of 100 pulsars in \autoref{fig:psr-distributions}, but the weights are arranged so that the first 10 have $w_i =1$, and the other 90 have $w_i = 0.05$.  This simulates having a small array of `good' pulsars, with many `bad' pulsars added to it.  The weights were chosen to make the effects in \autoref{fig:Kllmm} and \autoref{fig:svd} visible. 

Comparing the middle row of \autoref{fig:Kllmm} to the top and bottom shows that the coupling matrix is more similar to the case with 10 equal pulsars than to the case with 100 equal pulsars.  We can show this from \autoref{eq:coupling_matrix_sum}, assuming a simple distribution of weights, as in our example.

Consider two populations of pulsars: $N_g$ `good' pulsars with associated weights $w_g$ and coupling matrix $K_{(\ell m)(\ell m)'}^g$ and $N_b$ `bad' pulsars with associated weights $w_b$ and coupling matrix $K_{(\ell m)(\ell m)'}^b$. Since there are more `bad' pulsars than good ones, $K_{(\ell m)(\ell m)'}^b$ will typically have stronger diagonal terms, and weaker off-diagonal terms than $K_{(\ell m)(\ell m)'}^g$, although this will depend on the sky distribution of both sets of pulsars.

The contribution of each of these populations to the coupling matrix of the total population (from  \autoref{eq:coupling_matrix_sum}) can be separated:
\begin{align}
K_{(\ell m)(\ell m)'}^{g+b} = \frac{w_g N_g  K_{(\ell m)(\ell m)'}^g 
                                                + w_b N_b K_{(\ell m)(\ell m)'}^b}{w_g N_g+ w_b N_b}.
\label{eq:Kmix}
\end{align}
The overall coupling matrix is a weighted average between the coupling matrices calculated for each population on its own. Unless $N_b/N_g > w_g/w_b$,  `good' pulsars will be more important for the overall form of the coupling matrix. Real \acronym{pta}s tend to have sensitivities which are dominated by a small number of pulsars \citep[e.g.][where 6 out of 41 pulsars contribute more than 90\% of the total array's signal-to-noise squared]{babak16}, so similar results are likely to hold in practice.

Although the form of the coupling matrix is mostly set by the sub-array of `good' pulsars, adding a spatially extended distribution of `bad' pulsars can have a significant  effect on the conditioning of the coupling matrix (see \autoref{fig:svd}).  The best eigenvalues of $K_{(\ell m)(\ell m)'}^{g+b}$ are very similar to the best eigenvalues of $K_{(\ell m)(\ell m)'}^{g}$, but the worst are considerably improved.  This suggests that a large number of `bad' pulsars could be used to compensate for not having enough `good' pulsars to measure the desired $\ell_\text{max}$.  

\autoref{eq:Kmix} and Weyl's theorem can be used to place bounds on the amplitude of the smallest eigenvalue $\lambda_\text{min}$ for the combined arrays of `good' and `bad' pulsars:
\begin{align}
    \lambda_\text{min}^{g+b} &\geq  \frac{w_g N_g  \lambda_\text{min}^g 
                                                           + w_b N_b \lambda_\text{min}^b}
                                                            {w_g N_g+ w_b N_b} \\
     \lambda_\text{min}^{g+b} &\leq  \frac{w_g N_g  \lambda_\text{min}^g 
                                                               + w_b N_b \lambda_\text{max}^b}
                                                               {w_g N_g+ w_b N_b}.
\end{align}
This can be rewritten in terms of the improvement in $\lambda_\text{min}$ between the `good' array and the combined array:
\begin{equation}
F_b \left( \lambda_\text{min}^b - \lambda_\text{min}^g \right) \leq \Delta \lambda_\text{min} \leq F_b\left( \lambda_\text{max}^b - \lambda_\text{min}^g \right)
\end{equation}
where
$
F_b =  w_b N_b/(w_b N_b + w_g N_g).
$
 
The difference between the two bounds is related to the spatial distributions of the `good' and the `bad' pulsars.  The worst improvement $\propto \lambda_\text{min}^b - \lambda_\text{min}^g$ will occur when the two arrays have very similar distributions, and so make degenerate measurements.  In this case $\lambda_\text{min}^b$ may be similar in amplitude to $\lambda_\text{min}^g$, so the improvement can be very small.  

By contrast, the best improvement $\propto \lambda_\text{max}^b - \lambda_\text{min}^g$ will occur when the two arrays have complementary distributions, so that adding the `bad' pulsars add new information.  In this case, since $\lambda_\text{max}^b \gtrsim 1$, even a relatively small array of low-weight pulsars can have a significant improvement.

Although adding `bad' pulsars to the `good' ones typically will not significantly change the amount of coupling, they are most likely to be useful when they have a complementary sky distribution to the `good' ones. In this case the off-diagonal terms of their coupling matrices will partially cancel.

In summary, second-rate pulsars are generally less useful for making the sky distribution more uniform, since they have a weaker effect on the coupling matrix, but they are useful for improving the conditioning of the coupling matrix.  The best effect will be produced by a collection of `bad' pulsars with nearly the opposite distribution on the sky from the `good' pulsars, in enough numbers that $\sum w_b \sim \sum w_g $.

\section{Correcting for mode coupling}
\label{sec:dealing_with_Kllmm}

Large numbers of uniformly-distributed pulsars have very little coupling between the \acronym{gw} quadrupole (and higher) terms and spatially-correlated clock and ephemeris errors (monopole and dipole terms). However, since millisecond pulsars do not have a uniform sky distribution, in practice it may not be possible to construct a sensitive  \acronym{pta} without this coupling.

It is therefore useful to have techniques for removing the coupling.  
The standard  analysis uses the Hellings and Downs correlation between different pulsars to search for a \acronym{gwb}.  An equivalent harmonic space calculation could be made using the angular correlation in \autoref{eq:Cl}.

This is both optimal and unbiased if the monopole and dipole terms are both zero.  But any single attempted measurement of the \acronym{gwb} where either the monopole or dipole is nonzero will produce biased results \citep[cf.][]{tiburzi16,vigeland18} since power can leak from the monopole or dipole into the \acronym{gw} modes.

Knowledge of the coupling matrix can be used to explicitly remove the contamination, but care must be taken to avoid methods which are unbiased only for an ensemble of observations.  This section will explore a single example of how this can be done, but other techniques are possible \citep[cf.][]{tinto18}.

\subsection{Orthogonalized spherical harmonics}

One way to deal with the coupling is to build a new set of orthogonal basis functions based on the spherical harmonics.  As shown in \citet{gorski94a} and \citet{mortlock02} in the context of analysis of the cosmic microwave background anisotropies, this approach allows us to identify and excise components which are contaminated by the noisy monopole and dipole.  This section and the next will apply their strategy in the context of pulsar timing arrays.  For mathematical ease, the notation of the previous sections will be used interchangeably with linear algebra notation: $\mathbf{y} = Y_{\ell m}$.

The orthogonalization will be performed using a Cholesky decomposition of the coupling matrix:
\begin{equation}
    K_{(\ell m)(\ell m)'}  =  \mathbf{K} = \mathbf{L} \mathbf{L}^{\dagger},
\end{equation}
where $\mathbf{L}$ is a lower-triangular matrix, and $\mathbf{L}^\dagger$ represents the Hermitian transpose of $\mathbf{L}$.  

The new orthogonalized harmonics are given by:
\begin{equation}
\mathbf{y}' = \mathbf{L}^{-1}\mathbf{y}.
\label{eq:ortho-def}
\end{equation}
By construction, they are orthonormal on the space of the pulsar sky positions:
\begin{align}
    \int_\text{psrs} \mspace{-2mu} W(\Omega) \, \mathbf{y}'(\mathbf{y}')^\dagger \mathrm{d}\Omega 
    & =  \mathbf{L}^{-1} \left[\int_\text{psrs} \mspace{-2mu} W(\Omega) \, \mathbf{y}  \mathbf{y}^\dagger \mathrm{d}\Omega\right] (\mathbf{L}^{-1})^\dagger \nonumber \\
    & = \mathbf{L}^{-1}  \mathbf{K} (\mathbf{L}^{-1})^\dagger \nonumber \\
    & = \mathbf{L}^{-1}  \mathbf{L} \mathbf{L}^{\dagger} (\mathbf{L}^\dagger)^{-1} \nonumber \\
    & = \mathbb{I}.
\end{align}
The new coefficients $\mathbf{a} '$ are similarly related to the true harmonic coefficients:
\begin{align}
    \label{eq:ortho-coeff}
\mathbf{a} ' &= \mathbf{L}^{\dagger} \mathbf{a}
\end{align}
and can be estimated from the measured timing residuals (cf.\ \autoref{eq:alm-def}):
\begin{align}
    \mathbf{a} ' &= \int_\text{psrs} r(\hat{p}) \left[\mathbf{y}'(\hat{p})\right]^* \mathrm{d}\Omega. 
\end{align}

The spherical harmonics $\mathbf{y}$ are best described with two indices $(\ell, m)$, but the transformation in \autoref{eq:ortho-def} doesn't preserve these symmetries.  As a result, the orthogonalized spherical harmonics $\mathbf{y}'$ have only a single index (denoted hereafter by $b$ or $c$), which is related to the $\ell$ components of $\mathbf{y}$.

Since the transformation in \autoref{eq:ortho-coeff} is based on a Cholesky decomposition,  $\mathbf{L}^\dagger$ is upper-triangular.  This means that the components of $\mathbf{a}'$ contain contributions from a decreasing number of modes of $\mathbf{a}$.

The first term of $\mathbf{a}'$, $a_0'$, is made up of contributions from all $a_{\ell m}$. If the components of $ K_{(\ell m)(\ell m)'}$ are arranged in ascending order of the $\ell$'s, then $a_1'$ will contain contributions from all the $a_{\ell m}$ except for the monopole. Similarly, the next three terms of $\mathbf{a}'$ progressively drop the dipole terms.  Therefore, \emph{only} the first four components of $\mathbf{a} '$ contain any contribution from the monopole and dipole modes

\begin{figure}
    \includegraphics[trim=5pt 5pt 5pt 5pt, clip=true]{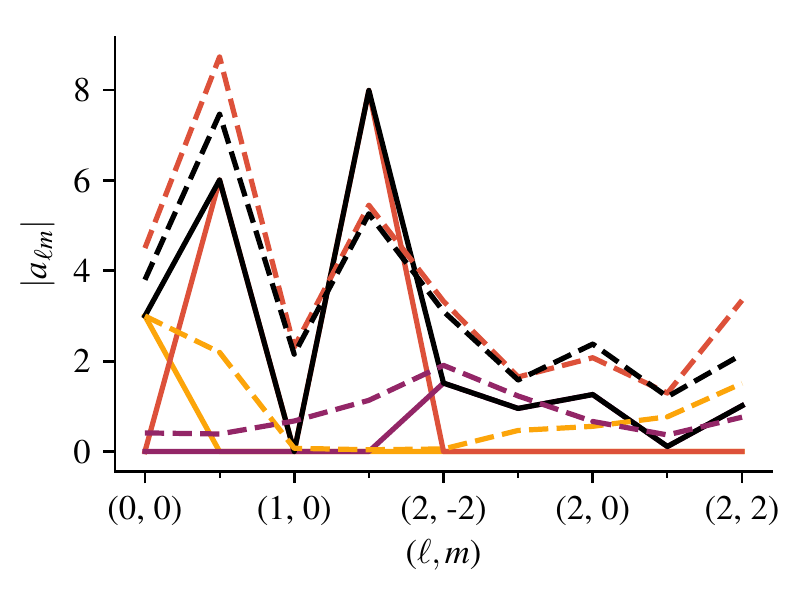}
    \includegraphics[trim=5pt 5pt 5pt 5pt, clip=true]{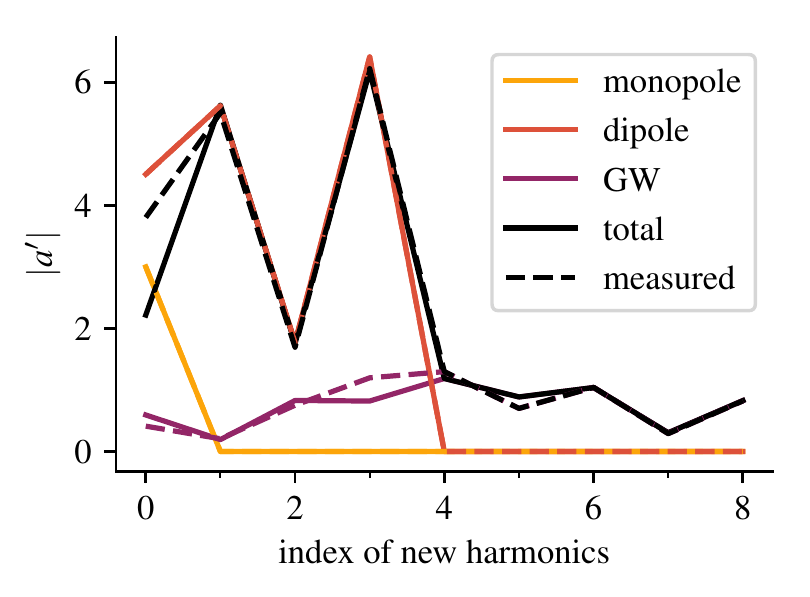}
    \caption{
        The measured and true amplitudes of the standard and new harmonics, calculated for a stochastic \acronym{gwb} (with the standard normalization), a monopole (amplitude 3), a dipole (amplitude 10) pointed in a random direction along the ecliptic, and the sum of all three fields.  For visual clarity, no pulsar noise is included.  Solid lines indicate true values and dashed lines indicate the amplitudes measured by an array with 100 equal galaxy-distributed pulsars.  The top panel shows the leakage of the monopole and dipole into the quadrupole terms. The new harmonics do not allow any leakage, but some of the power in the true quadrupole is lost. }
    \label{fig:alm-anew}
\end{figure}

By dropping those first four terms, we can cleanly separate the \acronym{gw} signal from the effect of the ephemeris and clock errors (see \autoref{fig:alm-anew}). Since the ambiguous components are also dropped, some amount of signal will be lost.  This can be calculated (on average) from the form of $\mathbf{L}^\dagger$.  In general, coupling matrices with large coupling terms and with larger condition numbers will lose more signal.

This process can be visualized as a switch to a new basis of functions which are built of spherical harmonics.  This new basis consists of (1) functions which contain the well-measured components of the quadrupole and (2) functions which contain the poorly-measured components of the quadrupole plus the monopole and dipole terms.  For an array with a lot of coupling, functions of type (1) will not cover much of the sky, and so a significant amount of signal will be lost.  By contrast, for an array with a nearly diagonal coupling matrix, functions of type (1) will approximately recover the quadrupole spherical harmonics, and very little signal will be lost.

Note that in ecliptic coordinates, we can expect that the ephemeris error will only contaminate the $m=\pm1$ modes, so we could limit ourselves to removing the contributions from only three modes: $Y_{00}$, $Y_{11}$, and $Y_{1,-1}$.  This would reduce the number of ambiguous modes, so that more signal could be recovered.

This technique is related to the one described by \cite{tinto18}, but leads to a very different set of uncontaminated modes, with very different covariance properties.  There are also other techniques that can be used to construct orthogonal modes from spherical harmonics.  The Cholesky decomposition was used for simplicity in constructing the clean modes $\mathbf{a}'$, but there are many other options.   \cite{mortlock02} discusses this problem more generally, including the case of a singular value decomposition, as well as techniques for extending the analysis to higher $\ell$, where the coupling matrix becomes singular.

\subsection{Likelihood analysis}

The vector $\mathbf{a}'$ can be thought of as a set of five linear combinations of the pulsar timing residuals, each of which is insensitive to the noisy monopole and dipole terms (note that we have now dropped the four contaminated modes).  In this section, we will discuss the statistical properties of  $\mathbf{a}'$ which are needed for \acronym{gwb} searches.  

For a Gaussian \acronym{gwb}, the spherical harmonic amplitudes $a_{\ell m}$ are by definition normally distributed.  The amplitudes $a'$ of the orthogonalized spherical harmonics can each be written as a sum of $a_{\ell m}$ components.  Therefore, the new amplitudes are also Gaussian random variables:
\begin{equation}
    P(\mathbf{a} ') = 
    \frac{\exp\left[-\frac{1}{2} (\mathbf{a} ')^\dagger 
            \mathbf{C}^{-1}\mathbf{a} '\right]}{\sqrt{(2\pi)^N\det\mathbf{C}}},
\end{equation}
where $N = (\ell_\text{max} + 1)^2 - N_\text{removed}$ is the number of $\mathbf{a} '$ modes.  If $\ell_\text{max}=2$ and all monopole and dipole modes are removed, $N=5$. 
The covariance matrix $\mathbf{C}$ can be separated into a signal component and pulsar noise component: $\mathbf{C} = \mathbf{C}_\textsc{gwb} + \mathbf{C}_N$.  

The signal covariance matrix can be defined in terms of the coefficients of the orthogonalized harmonics:
\begin{align}
    \mathbf{C}_\textsc{gwb} &  = \left\langle \mathbf{a}'  (\mathbf{a}')^\dag \right\rangle = \mathbf{L}^\dag \left\langle \mathbf{a}  \mathbf{a} ^\dag \right\rangle \mathbf{L} \nonumber \\
    & = \sum_{(\ell m)} L_{a (\ell m)}^* \, C_\ell L_{b(\ell m)},
    \label{eq:cov_GWB}
\end{align}
where $a$ and $b$ now represent the index of the orthogonal modes where all monopole and dipole terms have been dropped.
The second line is found using \autoref{eq:lm2pt} and noting that $\langle a_{\ell m} \rangle$ is the same for all $m$ at fixed $\ell$.  

In a full-sky analysis, the orthogonality of the spherical harmonics means that $\mathbf{C}_\textsc{gwb}$ is diagonal (with amplitudes given by $C_\ell$).  However, this is not generally the case for the orthogonalized spherical harmonics. As the array becomes larger and more uniform, the coupling matrix $\mathbf{K}$ will approach the identity matrix, and $\mathbf{C}_\textsc{gwb}$ will become approximately uniform.

Assuming that the noise in each pulsar is given by $\sigma_i$ and is uncorrelated with other pulsars, the noise covariance matrix can be written:
\begin{align} 
    C_N &= \langle \mathbf{n} \mathbf{n}^\dag \rangle =
     \frac{4\pi}{\sum_i w_i} \left \langle (\sigma_i w_i \mathbf{y}'_i) (\sigma_j w_j \mathbf{y}'_j)^\dag \delta_{ij} \right \rangle \\
     & = \frac{4\pi}{\sum_i w_i} \left \langle \sum_i \sigma_i^2 w_i^2 \mathbf{y}'_i (\mathbf{y}'_i)^\dag \right \rangle
     \label{eq:cov_N}
\end{align}
This is diagonal and exactly equal to the identity matrix if all pulsars are identical, or if the pulsar weights are equal to the inverse variance of their noise: $w_i = 1/\sigma_i^2$.

\section{Discussion}
\label{sec:conclusion}

The form of the antenna response function ensures that \acronym{pta} searches for \acronym{gw}s have an inherently geometric component.  This is fortunate, since it gives us a way to decouple the \acronym{gw} signal from correlated noise sources such as planetary ephemeris errors and clock errors.  

A natural space in which to do this is provided by a spherical harmonic analysis of the pulsar timing residuals.  In this space, \acronym{gw}s appear as quadrupole and higher order terms, while ephemeris errors appear as dipole terms, and clock errors appear as a monopole.  

However, real \acronym{pta}s are composed of a moderate number of pulsars with varying intrinsic noise levels, distributed nonuniformly across the sky.  Attempts to measure the different modes are therefore subject to aliasing-type effects, and the measured quadrupole \acronym{gw} signal may be contaminated by the clock error monopole and ephemeris error dipole. 

This contamination can be quantified in terms of a coupling matrix, which measures the extent to which different spherical harmonics are non-orthogonal for a given array of pulsars.  The coupling matrix depends only on the spatial arrangement and noise properties of the pulsars in the array, and so provides a means to assess the degree to which contamination by ephemeris errors or clock errors is likely to be an issue for a particular configuration of pulsars.  

Measuring the monopole, dipole, and quadrupole modes requires an array with a minimum of 9 pulsars, but such arrays are typically subject to considerable coupling between the \acronym{gw} signal and ephemeris or clock errors.  The arrays with the fewest coupling problems are those with many isotropic, high-quality pulsars.  However, since the underlying distribution of millisecond pulsars is anisotropic, future \acronym{pta}s will probably still experience significant amounts of coupling.  

A biased array can be \replaced{improved}{made less biased} by keeping only the best pulsars from the region around galactic center and by adding clusters of second-rate pulsars wherever possible in the undersampled regions.  In particular, covering the ecliptic plane is especially useful for minimizing coupling with the dipole modes most likely to be induced by ephemeris errors.  

The distribution of pulsars which allows the \acronym{gw} signal to be most cleanly separated from the clock and ephemeris errors is not the same as the one which is most sensitive to detecting \acronym{gw}s, since the latter case favors closely-spaced pulsars, which tend to increase coupling.  \added{Moreover, choosing pulsars such that the sky coverage is as uniform as possible will lead to a loss of sensitivity to \acronym{gw}s if good pulsars are dropped.  }

\added{An array which can make a clean detection of \acronym{gw}s will need to balance these two conflicting guidelines, by ensuring that all pulsars are evaluated both in terms of their own quality and also in terms of their effect on the coupling matrix.  As long as a detection has not yet been made, increasing the sensitivity of the array will remain the priority.  However, improving the sky coverage can significantly improve the conditioning of the coupling matrix, even if only second-rate pulsars are used.  Such arrays will still exhibit significant amounts of coupling, but should allow the \acronym{gw} signal to be disentangled from the effects of ephemeris and clock errors. }

A \acronym{pta} with significant coupling will need to correct for that coupling.  If the array contains a large enough number of pulsars, the coupling matrix will be invertible, and the coupling can be removed algebraically.  \added{Making a clean cut between the coupling and the signal will necessitate some loss of signal.  More signal will be retained when the coupling matrix is well-conditioned and has small off-diagonal components.}  This paper and \cite{tinto18} present different examples of how the analysis can be done in a space which nulls contributions from the monopole and dipole terms.  These algebraic techniques are complementary to physical models such as the planetary ephemeris model \textsc{BayesEphem} \citep{arzoumanian18}, \added{but it is likely that any analysis will benefit if pulsar arrays are chosen to minimize coupling while maintaining sensitivity to \acronym{gw}s}. 

Throughout this paper, we have assumed that ephemeris and clock errors produce pure monopolar and dipolar correlations between all pulsars in the array, but this will be more complicated in reality.  In particular, timing residuals from different pulsars are irregularly sampled, there are gaps in the timing of certain pulsars, and new pulsars are occasionally added to the array.  As a result, a real \acronym{pta} cannot be treated as a static collection of pulsars, and a frequency-domain analysis becomes difficult.  

\added{For this reason, \acronym{pta} analyses are typically performed in the time domain.}
It is possible to extend the analysis of this paper \replaced{to treat this}{to the time domain}.  The time-domain coupling matrix can be calculated with \autoref{eq:coupling_matrix_sum}, using real spherical harmonics and time-varying pulsar positions and weights.   
The single coupling matrix that we have been considering becomes instead a series of different coupling matrices as pulsars are added and subtracted.  As a result, the degree of contamination by ephemeris and clock errors will also vary as a function of time.  If the analysis above is followed, this will also lead to a choice of  $\mathbf{y}'(t)$ which is not smoothly varying.  
\added{However, since this analysis requires simultaneously combining data from all the pulsars in the array, rather than two at a time as in the standard analysis, irregular sampling may make this difficult to implement. }  

\added{The full time-domain Bayesian search for \acronym{gw}s is extremely computationally expensive, since the combination of timing residuals from all pulsars results in a very large covariance matrix.  This covariance matrix cannot be precomputed, since it depends on the model parameters, and must be inverted for each likelihood evaluation.  In search of a more computationally efficient analysis, recent work often treats the model parameters for the correlated components with a reduced-rank analysis which, through the application of known transformations, approximates the full time-domain covariance matrix \citep{lentati13,van-haasteren14,lentati15,arzoumanian18}.  This is typically an approximate Fourier decomposition, in which the low-frequency components (which contain most of the information) are retained \citep[although see also][]{van-haasteren15}.} 

\added{The pulsar correlations due to \acronym{gw}s are treated in this stage of the analysis, through our knowledge of their expected frequency and spatial correlations.  The analysis presented in this work could potentially be used in the place of the traditional Hellings and Downs correlations to treat the spatial part.  This would lead to a new transformation between the orthogonalized spherical harmonics and the individual pulsar correlations, similar to the transformation between the frequency-domain and the time-domain correlations.  These transformations would probably need to be done with careful treatment of pulsar weights at the different frequencies to avoid reintroducing monopole or dipole contamination via spectral leakage.}  

\added{Implementing the techniques described in this paper within the reduced-rank variant of the full time-domain Gaussian analysis overcomes some of the difficulties presented by irregular sampling, since all pulsars can be transformed to a consistent set of frequencies.  It could also have the benefit of further reducing the rank of the modeled covariance matrix from $ N_\text{freq} \times N_\text{psr}$ to $N_\text{freq} \times 5$, where $N_\text{freq}$ is the number of frequency components considered. This reduction comes from the fact that the spatial correlations can be described using the $5\times5$ orthogonalized spherical harmonic covariance matrix from \autoref{eq:cov_GWB} and \autoref{eq:cov_N}, rather than the standard $N_\text{psr} \times N_\text{psr}$ covariance matrix. }

\replaced{This analysis has}{The analysis presented in this paper has also} treated only the case where the \acronym{gwb} is an isotropic Gaussian random field.  This may be an oversimplification.
It is plausible that strong individual \acronym{gw} sources will dominate the signal in some frequency bins \citep{sesana08,ravi12a,roebber16}.  Although there are better means to detect single sources \citep[e.g.][]{lee11,boyle12,babak12,zhu16,goldstein18}, their presence should not significantly bias our analysis \citep[cf.][]{cornish16}.

 \replaced{ Single sources have the same expected harmonic power spectrum as an isotropic \acronym{gwb}, but with correlations between different $\ell$   \mbox{\citep{roebber17}}. }{Single sources of \acronym{gw}s also produce Hellings and Downs correlations between pulsars, and, equivalently, the same angular power spectrum as in \autoref{eq:Cl}.  For multiple sources, as with individual realizations of the stochastic \acronym{gwb}, this correlation represents an ensemble average and does not hold exactly \mbox{\citep{roebber17}}.}  
 
 \replaced{Since these correlations are not typically larger than the intrinsic variations, they should be unimportant while \acronym{pta}s are only sensitive to the quadrupole term.}{The primary difference between these cases is that for a stochastic \acronym{gwb}, the power in each multipole is random and uncorrelated, while for a \acronym{gw} signal produced by a few loud sources, the power in different multipoles becomes correlated due to interference between the pulsar antenna response to each source.  These correlations appear as small ripples in the angular power spectrum \mbox{\citep{roebber17}}. Since the underlying form of the angular power spectrum is similar in all cases, it will almost always be appropriate to approximate it as a quadrupole.  As a result, the technique described in this paper should work even when our assumptions of a statistically isotropic and Gaussian random \acronym{gwb} is violated by the presence of a few loud sources.}

Large-scale anisotropies in the distribution of \acronym{gw} sources \citep{mingarelli13,taylor13}, however, could produce a \acronym{gwb} with a very different \added{angular} power spectrum.  In this case, the assumption that the \acronym{gw} signal is largely contained in the quadrupole modes will no longer hold.  However, this scenario is unlikely in practice, as it represents a significant departure from cosmological isotropy.

The problems discussed in this paper will also affect astrometric \acronym{gw} searches using Gaia and similar experiments \citep{klioner18,obeirne18,darling18,mihaylov18}.  Gaia observes many stars, but star locations are biased towards the direction of the galaxy, so  similar coupling problems will arise.

Similarly, measurements of just the clock \citep{hobbs12} or the planetary ephemeris \citep{champion10} errors could experience significant coupling with the other effect or with \acronym{gw}s.  A variant of the method described here may be useful for attempts to measure these individual quantities.

Although the problem of ephemeris errors is not trivial, it is solvable, especially for future arrays with many high-quality pulsars.

\acknowledgements{I thank Alberto Vecchio and Alberto Sesana for useful comments.}

\software{Python with the following packages: numpy \citep{numpy}, scipy \citep{scipy}, pandas \citep{pandas}, healpy \citep{gorski05}, matplotlib \citep{matplotlib}, seaborn \citep{seaborn}.}

\bibliography{pta-ephemeris-mode-coupling}

\begin{thebibliography}{63}
\providecommand\natexlab[1]{#1}
\providecommand\JournalTitle[1]{#1}

\bibitem[{{Arzoumanian} {et~al.}(2018){Arzoumanian}, {Baker}, {Brazier},
  {Burke-Spolaor}, {Chamberlin}, {Chatterjee}, {Christy}, {Cordes}, {Cornish},
  {Crawford}, {Thankful Cromartie}, {Crowter}, {DeCesar}, {Demorest}, {Dolch},
  {Ellis}, {Ferdman}, {Ferrara}, {Folkner}, {Fonseca}, {Garver-Daniels},
  {Gentile}, {Haas}, {Hazboun}, {Huerta}, {Islo}, {Jones}, {Jones}, {Kaplan},
  {Kaspi}, {Lam}, {Lazio}, {Levin}, {Lommen}, {Lorimer}, {Luo}, {Lynch},
  {Madison}, {McLaughlin}, {McWilliams}, {Mingarelli}, {Ng}, {Nice}, {Park},
  {Pennucci}, {Pol}, {Ransom}, {Ray}, {Rasskazov}, {Siemens}, {Simon},
  {Spiewak}, {Stairs}, {Stinebring}, {Stovall}, {Swiggum}, {Taylor},
  {Vallisneri}, {van Haasteren}, {Vigeland}, {Zhu}, \& {The NANOGrav
  Collaboration}}]{arzoumanian18}
{Arzoumanian}, Z., {Baker}, P.~T., {Brazier}, A., {et~al.} 2018,
  \href{http://dx.doi.org/10.3847/1538-4357/aabd3b}{\JournalTitle{\apj}, 859,
  47}

\bibitem[{{Babak} \& {Sesana}(2012)}]{babak12}
{Babak}, S., \& {Sesana}, A. 2012,
  \href{http://dx.doi.org/10.1103/PhysRevD.85.044034}{\JournalTitle{\prd}, 85,
  044034}

\bibitem[{{Babak} {et~al.}(2016){Babak}, {Petiteau}, {Sesana}, {Brem},
  {Rosado}, {Taylor}, {Lassus}, {Hessels}, {Bassa}, {Burgay}, {Caballero},
  {Champion}, {Cognard}, {Desvignes}, {Gair}, {Guillemot}, {Janssen},
  {Karuppusamy}, {Kramer}, {Lazarus}, {Lee}, {Lentati}, {Liu}, {Mingarelli},
  {Os{\l}owski}, {Perrodin}, {Possenti}, {Purver}, {Sanidas}, {Smits},
  {Stappers}, {Theureau}, {Tiburzi}, {van Haasteren}, {Vecchio}, \&
  {Verbiest}}]{babak16}
{Babak}, S., {Petiteau}, A., {Sesana}, A., {et~al.} 2016,
  \href{http://dx.doi.org/10.1093/mnras/stv2092}{\JournalTitle{\mnras}, 455,
  1665}

\bibitem[{{Begelman} {et~al.}(1980){Begelman}, {Blandford}, \&
  {Rees}}]{begelman80}
{Begelman}, M.~C., {Blandford}, R.~D., \& {Rees}, M.~J. 1980,
  \href{http://dx.doi.org/10.1038/287307a0}{\JournalTitle{\nat}, 287, 307}

\bibitem[{{Boyle} \& {Pen}(2012)}]{boyle12}
{Boyle}, L., \& {Pen}, U.-L. 2012,
  \href{http://dx.doi.org/10.1103/PhysRevD.86.124028}{\JournalTitle{\prd}, 86,
  124028}

\bibitem[{{Champion} {et~al.}(2010){Champion}, {Hobbs}, {Manchester},
  {Edwards}, {Backer}, {Bailes}, {Bhat}, {Burke-Spolaor}, {Coles}, {Demorest},
  {Ferdman}, {Folkner}, {Hotan}, {Kramer}, {Lommen}, {Nice}, {Purver},
  {Sarkissian}, {Stairs}, {van Straten}, {Verbiest}, \& {Yardley}}]{champion10}
{Champion}, D.~J., {Hobbs}, G.~B., {Manchester}, R.~N., {et~al.} 2010,
  \href{http://dx.doi.org/10.1088/2041-8205/720/2/L201}{\JournalTitle{\apjl},
  720, L201}

\bibitem[{{Chen} {et~al.}(2018){Chen}, {Sesana}, \& {Conselice}}]{chen18a}
{Chen}, S., {Sesana}, A., \& {Conselice}, C.~J. 2018,
  \href{http://arxiv.org/abs/1810.04184}{{\sffamily arXiv:1810.04184}}

\bibitem[{{Cordes}(2013)}]{cordes13}
{Cordes}, J.~M. 2013,
  \href{http://dx.doi.org/10.1088/0264-9381/30/22/224002}{\JournalTitle{Classical
  and Quantum Gravity}, 30, 224002}

\bibitem[{{Cornish} \& {Sampson}(2016)}]{cornish16}
{Cornish}, N.~J., \& {Sampson}, L. 2016,
  \href{http://dx.doi.org/10.1103/PhysRevD.93.104047}{\JournalTitle{\prd}, 93,
  104047}

\bibitem[{{Darling} {et~al.}(2018){Darling}, {Truebenbach}, \&
  {Paine}}]{darling18}
{Darling}, J., {Truebenbach}, A.~E., \& {Paine}, J. 2018,
  \href{http://dx.doi.org/10.3847/1538-4357/aac772}{\JournalTitle{\apj}, 861,
  113}

\bibitem[{{Dodelson}(2003)}]{dodelson03}
{Dodelson}, S. 2003, {Modern cosmology} (Academic Press)

\bibitem[{{Edwards} {et~al.}(2006){Edwards}, {Hobbs}, \&
  {Manchester}}]{edwards06}
{Edwards}, R.~T., {Hobbs}, G.~B., \& {Manchester}, R.~N. 2006,
  \href{http://dx.doi.org/10.1111/j.1365-2966.2006.10870.x}{\JournalTitle{\mnras},
  372, 1549}

\bibitem[{{Efstathiou}(2004)}]{efstathiou04}
{Efstathiou}, G. 2004,
  \href{http://dx.doi.org/10.1111/j.1365-2966.2004.07530.x}{\JournalTitle{\mnras},
  349, 603}

\bibitem[{{Foster} \& {Backer}(1990)}]{foster90}
{Foster}, R.~S., \& {Backer}, D.~C. 1990,
  \href{http://dx.doi.org/10.1086/169195}{\JournalTitle{\apj}, 361, 300}

\bibitem[{{Gair} {et~al.}(2014){Gair}, {Romano}, {Taylor}, \&
  {Mingarelli}}]{gair14}
{Gair}, J., {Romano}, J.~D., {Taylor}, S., \& {Mingarelli}, C.~M.~F. 2014,
  \href{http://dx.doi.org/10.1103/PhysRevD.90.082001}{\JournalTitle{\prd}, 90,
  082001}

\bibitem[{{Goldstein} {et~al.}(2018){Goldstein}, {Veitch}, {Sesana}, \&
  {Vecchio}}]{goldstein18}
{Goldstein}, J.~M., {Veitch}, J., {Sesana}, A., \& {Vecchio}, A. 2018,
  \href{http://dx.doi.org/10.1093/mnras/sty892}{\JournalTitle{\mnras}, 477,
  5447}

\bibitem[{{Gorski}(1994)}]{gorski94a}
{Gorski}, K.~M. 1994,
  \href{http://dx.doi.org/10.1086/187444}{\JournalTitle{\apjl}, 430, L85}

\bibitem[{{G{\'o}rski} {et~al.}(2005){G{\'o}rski}, {Hivon}, {Banday},
  {Wandelt}, {Hansen}, {Reinecke}, \& {Bartelmann}}]{gorski05}
{G{\'o}rski}, K.~M., {Hivon}, E., {Banday}, A.~J., {et~al.} 2005,
  \href{http://dx.doi.org/10.1086/427976}{\JournalTitle{\apj}, 622, 759}

\bibitem[{{Hellings} \& {Downs}(1983)}]{hellings83}
{Hellings}, R.~W., \& {Downs}, G.~S. 1983,
  \href{http://dx.doi.org/10.1086/183954}{\JournalTitle{\apjl}, 265, L39}

\bibitem[{{Hivon} {et~al.}(2002){Hivon}, {G{\'o}rski}, {Netterfield}, {Crill},
  {Prunet}, \& {Hansen}}]{hivon02}
{Hivon}, E., {G{\'o}rski}, K.~M., {Netterfield}, C.~B., {et~al.} 2002,
  \href{http://dx.doi.org/10.1086/338126}{\JournalTitle{\apj}, 567, 2}

\bibitem[{{Hobbs} \& {Dai}(2017)}]{hobbs17}
{Hobbs}, G., \& {Dai}, S. 2017,
  \href{http://arxiv.org/abs/1707.01615}{{\sffamily arXiv:1707.01615}}

\bibitem[{{Hobbs} {et~al.}(2012){Hobbs}, {Coles}, {Manchester}, {Keith},
  {Shannon}, {Chen}, {Bailes}, {Bhat}, {Burke-Spolaor}, {Champion},
  {Chaudhary}, {Hotan}, {Khoo}, {Kocz}, {Levin}, {Oslowski}, {Preisig}, {Ravi},
  {Reynolds}, {Sarkissian}, {van Straten}, {Verbiest}, {Yardley}, \&
  {You}}]{hobbs12}
{Hobbs}, G., {Coles}, W., {Manchester}, R.~N., {et~al.} 2012,
  \href{http://dx.doi.org/10.1111/j.1365-2966.2012.21946.x}{\JournalTitle{\mnras},
  427, 2780}

\bibitem[{Hunter(2007)}]{matplotlib}
Hunter, J.~D. 2007,
  \href{http://dx.doi.org/10.1109/MCSE.2007.55}{\JournalTitle{Computing In
  Science \& Engineering}, 9, 90}

\bibitem[{Jones {et~al.}(2001--present)Jones, Oliphant, Peterson, \&
  et~al.}]{scipy}
Jones, E., Oliphant, T., Peterson, P., \& et~al. 2001--present, {SciPy}: Open
  source scientific tools for {Python}

\bibitem[{{Joshi} {et~al.}(2018){Joshi}, {Arumugasamy}, {Bagchi},
  {Bandyopadhyay}, {Basu}, {Dhanda Batra}, {Bethapudi}, {Choudhary}, {De},
  {Dey}, {Gopakumar}, {Gupta}, {Krishnakumar}, {Maan}, {Manoharan}, {Naidu},
  {Nandi}, {Pathak}, {Surnis}, \& {Susobhanan}}]{joshi18}
{Joshi}, B.~C., {Arumugasamy}, P., {Bagchi}, M., {et~al.} 2018,
  \href{http://dx.doi.org/10.1007/s12036-018-9549-y}{\JournalTitle{Journal of
  Astrophysics and Astronomy}, 39, 51}

\bibitem[{{Keane} {et~al.}(2015){Keane}, {Bhattacharyya}, {Kramer}, {Stappers},
  {Keane}, {Bhattacharyya}, {Kramer}, {Stappers}, {Bates}, {Burgay},
  {Chatterjee}, {Champion}, {Eatough}, {Hessels}, {Janssen}, {Lee}, {van
  Leeuwen}, {Margueron}, {Oertel}, {Possenti}, {Ransom}, {Theureau}, \&
  {Torne}}]{keane15}
{Keane}, E., {Bhattacharyya}, B., {Kramer}, M., {et~al.} 2015, in Advancing
  Astrophysics with the Square Kilometre Array (AASKA14), 40

\bibitem[{{Klioner}(2018)}]{klioner18}
{Klioner}, S.~A. 2018,
  \href{http://dx.doi.org/10.1088/1361-6382/aa9f57}{\JournalTitle{Classical and
  Quantum Gravity}, 35, 045005}

\bibitem[{{Lasky} {et~al.}(2016){Lasky}, {Mingarelli}, {Smith}, {Giblin},
  {Thrane}, {Reardon}, {Caldwell}, {Bailes}, {Bhat}, {Burke-Spolaor}, {Dai},
  {Dempsey}, {Hobbs}, {Kerr}, {Levin}, {Manchester}, {Os{\l}owski}, {Ravi},
  {Rosado}, {Shannon}, {Spiewak}, {van Straten}, {Toomey}, {Wang}, {Wen},
  {You}, \& {Zhu}}]{lasky16}
{Lasky}, P.~D., {Mingarelli}, C.~M.~F., {Smith}, T.~L., {et~al.} 2016,
  \href{http://dx.doi.org/10.1103/PhysRevX.6.011035}{\JournalTitle{Physical
  Review X}, 6, 011035}

\bibitem[{{Lazio} {et~al.}(2018){Lazio}, {Bhaskaran}, {Cutler}, {Folkner},
  {Park}, {Ellis}, {Ely}, {Taylor}, \& {Vallisneri}}]{lazio18}
{Lazio}, T.~J.~W., {Bhaskaran}, S., {Cutler}, C., {et~al.} 2018,
  \href{http://dx.doi.org/10.1017/S1743921317009711}{in IAU Symposium, Vol.
  337, Pulsar Astrophysics the Next Fifty Years, ed. P.~{Weltevrede}, B.~B.~P.
  {Perera}, L.~L. {Preston}, \& S.~{Sanidas}}, 150

\bibitem[{{Lee}(2016)}]{lee16}
{Lee}, K.~J. 2016, in Astronomical Society of the Pacific Conference Series,
  Vol. 502, Frontiers in Radio Astronomy and FAST Early Sciences Symposium
  2015, ed. L.~{Qain} \& D.~{Li}, 19

\bibitem[{{Lee} {et~al.}(2011){Lee}, {Wex}, {Kramer}, {Stappers}, {Bassa},
  {Janssen}, {Karuppusamy}, \& {Smits}}]{lee11}
{Lee}, K.~J., {Wex}, N., {Kramer}, M., {et~al.} 2011,
  \href{http://dx.doi.org/10.1111/j.1365-2966.2011.18622.x}{\JournalTitle{\mnras},
  414, 3251}

\bibitem[{{Lentati} {et~al.}(2013){Lentati}, {Alexander}, {Hobson}, {Taylor},
  {Gair}, {Balan}, \& {van Haasteren}}]{lentati13}
{Lentati}, L., {Alexander}, P., {Hobson}, M.~P., {et~al.} 2013,
  \href{http://dx.doi.org/10.1103/PhysRevD.87.104021}{\JournalTitle{\prd}, 87,
  104021}

\bibitem[{{Lentati} {et~al.}(2015){Lentati}, {Taylor}, {Mingarelli}, {Sesana},
  {Sanidas}, {Vecchio}, {Caballero}, {Lee}, {van Haasteren}, {Babak}, {Bassa},
  {Brem}, {Burgay}, {Champion}, {Cognard}, {Desvignes}, {Gair}, {Guillemot},
  {Hessels}, {Janssen}, {Karuppusamy}, {Kramer}, {Lassus}, {Lazarus}, {Liu},
  {Os{\l}owski}, {Perrodin}, {Petiteau}, {Possenti}, {Purver}, {Rosado},
  {Smits}, {Stappers}, {Theureau}, {Tiburzi}, \& {Verbiest}}]{lentati15}
{Lentati}, L., {Taylor}, S.~R., {Mingarelli}, C.~M.~F., {et~al.} 2015,
  \href{http://dx.doi.org/10.1093/mnras/stv1538}{\JournalTitle{\mnras}, 453,
  2576}

\bibitem[{{Lorimer}(2008)}]{lorimer08}
{Lorimer}, D.~R. 2008,
  \href{http://dx.doi.org/10.12942/lrr-2008-8}{\JournalTitle{Living Reviews in
  Relativity}, 11}, \href{http://arxiv.org/abs/0811.0762}{{\sffamily
  arXiv:0811.0762}}

\bibitem[{{Manchester} {et~al.}(2005){Manchester}, {Hobbs}, {Teoh}, \&
  {Hobbs}}]{manchester05}
{Manchester}, R.~N., {Hobbs}, G.~B., {Teoh}, A., \& {Hobbs}, M. 2005,
  \href{http://dx.doi.org/10.1086/428488}{\JournalTitle{\aj}, 129, 1993}

\bibitem[{McKinney(2010)}]{pandas}
McKinney, W. 2010, in Proceedings of the 9th Python in Science Conference, ed.
  S.~van~der Walt \& J.~Millman, 51

\bibitem[{{Mihaylov} {et~al.}(2018){Mihaylov}, {Moore}, {Gair}, {Lasenby}, \&
  {Gilmore}}]{mihaylov18}
{Mihaylov}, D.~P., {Moore}, C.~J., {Gair}, J.~R., {Lasenby}, A., \& {Gilmore},
  G. 2018,
  \href{http://dx.doi.org/10.1103/PhysRevD.97.124058}{\JournalTitle{\prd}, 97,
  124058}

\bibitem[{{Mingarelli} {et~al.}(2013){Mingarelli}, {Sidery}, {Mandel}, \&
  {Vecchio}}]{mingarelli13}
{Mingarelli}, C.~M.~F., {Sidery}, T., {Mandel}, I., \& {Vecchio}, A. 2013,
  \href{http://dx.doi.org/10.1103/PhysRevD.88.062005}{\JournalTitle{\prd}, 88,
  062005}

\bibitem[{{Mortlock} {et~al.}(2002){Mortlock}, {Challinor}, \&
  {Hobson}}]{mortlock02}
{Mortlock}, D.~J., {Challinor}, A.~D., \& {Hobson}, M.~P. 2002,
  \href{http://dx.doi.org/10.1046/j.1365-8711.2002.05085.x}{\JournalTitle{\mnras},
  330, 405}

\bibitem[{{O'Beirne} \& {Cornish}(2018)}]{obeirne18}
{O'Beirne}, L., \& {Cornish}, N.~J. 2018,
  \href{http://dx.doi.org/10.1103/PhysRevD.98.024020}{\JournalTitle{\prd}, 98,
  024020}

\bibitem[{{Peebles}(1973)}]{peebles73}
{Peebles}, P.~J.~E. 1973,
  \href{http://dx.doi.org/10.1086/152431}{\JournalTitle{\apj}, 185, 413}

\bibitem[{{Rajagopal} \& {Romani}(1995)}]{rajagopal95}
{Rajagopal}, M., \& {Romani}, R.~W. 1995,
  \href{http://dx.doi.org/10.1086/175813}{\JournalTitle{\apj}, 446, 543}

\bibitem[{{Ravi} {et~al.}(2012){Ravi}, {Wyithe}, {Hobbs}, {Shannon},
  {Manchester}, {Yardley}, \& {Keith}}]{ravi12a}
{Ravi}, V., {Wyithe}, J.~S.~B., {Hobbs}, G., {et~al.} 2012,
  \href{http://dx.doi.org/10.1088/0004-637X/761/2/84}{\JournalTitle{\apj}, 761,
  84}

\bibitem[{{Roebber} \& {Holder}(2017)}]{roebber17}
{Roebber}, E., \& {Holder}, G. 2017,
  \href{http://dx.doi.org/10.3847/1538-4357/835/1/21}{\JournalTitle{\apj}, 835,
  21}

\bibitem[{{Roebber} {et~al.}(2016){Roebber}, {Holder}, {Holz}, \&
  {Warren}}]{roebber16}
{Roebber}, E., {Holder}, G., {Holz}, D.~E., \& {Warren}, M. 2016,
  \href{http://dx.doi.org/10.3847/0004-637X/819/2/163}{\JournalTitle{\apj},
  819, 163}

\bibitem[{{Romano} \& {Cornish}(2017)}]{romano17}
{Romano}, J.~D., \& {Cornish}, N.~J. 2017,
  \href{http://dx.doi.org/10.1007/s41114-017-0004-1}{\JournalTitle{Living
  Reviews in Relativity}, 20, 2}

\bibitem[{{Sesana} {et~al.}(2008){Sesana}, {Vecchio}, \& {Colacino}}]{sesana08}
{Sesana}, A., {Vecchio}, A., \& {Colacino}, C.~N. 2008,
  \href{http://dx.doi.org/10.1111/j.1365-2966.2008.13682.x}{\JournalTitle{\mnras},
  390, 192}

\bibitem[{{Shannon} {et~al.}(2015){Shannon}, {Ravi}, {Lentati}, {Lasky},
  {Hobbs}, {Kerr}, {Manchester}, {Coles}, {Levin}, {Bailes}, {Bhat},
  {Burke-Spolaor}, {Dai}, {Keith}, {Os{\l}owski}, {Reardon}, {van Straten},
  {Toomey}, {Wang}, {Wen}, {Wyithe}, \& {Zhu}}]{shannon15}
{Shannon}, R.~M., {Ravi}, V., {Lentati}, L.~T., {et~al.} 2015,
  \href{http://dx.doi.org/10.1126/science.aab1910}{\JournalTitle{Science}, 349,
  1522}

\bibitem[{{Siemens} {et~al.}(2013){Siemens}, {Ellis}, {Jenet}, \&
  {Romano}}]{siemens13}
{Siemens}, X., {Ellis}, J., {Jenet}, F., \& {Romano}, J.~D. 2013,
  \href{http://dx.doi.org/10.1088/0264-9381/30/22/224015}{\JournalTitle{Classical
  and Quantum Gravity}, 30, 224015}

\bibitem[{{Taylor} \& {Gair}(2013)}]{taylor13}
{Taylor}, S.~R., \& {Gair}, J.~R. 2013,
  \href{http://dx.doi.org/10.1103/PhysRevD.88.084001}{\JournalTitle{\prd}, 88,
  084001}

\bibitem[{{Taylor} {et~al.}(2017{\natexlab{a}}){Taylor}, {Lentati}, {Babak},
  {Brem}, {Gair}, {Sesana}, \& {Vecchio}}]{taylor17a}
{Taylor}, S.~R., {Lentati}, L., {Babak}, S., {et~al.} 2017{\natexlab{a}},
  \href{http://dx.doi.org/10.1103/PhysRevD.95.042002}{\JournalTitle{\prd}, 95,
  042002}

\bibitem[{{Taylor} {et~al.}(2017{\natexlab{b}}){Taylor}, {Simon}, \&
  {Sampson}}]{taylor17}
{Taylor}, S.~R., {Simon}, J., \& {Sampson}, L. 2017{\natexlab{b}},
  \href{http://dx.doi.org/10.1103/PhysRevLett.118.181102}{\JournalTitle{\prl},
  118, 181102}

\bibitem[{{Taylor} {et~al.}(2016){Taylor}, {Vallisneri}, {Ellis}, {Mingarelli},
  {Lazio}, \& {van Haasteren}}]{taylor16a}
{Taylor}, S.~R., {Vallisneri}, M., {Ellis}, J.~A., {et~al.} 2016,
  \href{http://dx.doi.org/10.3847/2041-8205/819/1/L6}{\JournalTitle{\apjl},
  819, L6}

\bibitem[{{Tiburzi} {et~al.}(2016){Tiburzi}, {Hobbs}, {Kerr}, {Coles}, {Dai},
  {Manchester}, {Possenti}, {Shannon}, \& {You}}]{tiburzi16}
{Tiburzi}, C., {Hobbs}, G., {Kerr}, M., {et~al.} 2016,
  \href{http://dx.doi.org/10.1093/mnras/stv2143}{\JournalTitle{\mnras}, 455,
  4339}

\bibitem[{{Tinto}(2018)}]{tinto18}
{Tinto}, M. 2018,
  \href{http://dx.doi.org/10.1103/PhysRevD.97.084047}{\JournalTitle{\prd}, 97,
  084047}

\bibitem[{{van der Walt} {et~al.}(2011){van der Walt}, Colbert, \&
  Varoquaux}]{numpy}
{van der Walt}, S., Colbert, S.~C., \& Varoquaux, G. 2011,
  \href{http://dx.doi.org/10.1109/MCSE.2011.37}{\JournalTitle{Computing in
  Science \& Engineering}, 13, 22}

\bibitem[{{van Haasteren} \& {Vallisneri}(2014)}]{van-haasteren14}
{van Haasteren}, R., \& {Vallisneri}, M. 2014,
  \href{http://dx.doi.org/10.1103/PhysRevD.90.104012}{\JournalTitle{\prd}, 90,
  104012}

\bibitem[{{van Haasteren} \& {Vallisneri}(2015)}]{van-haasteren15}
---. 2015,
  \href{http://dx.doi.org/10.1093/mnras/stu2157}{\JournalTitle{\mnras}, 446,
  1170}

\bibitem[{{Verbiest} {et~al.}(2016){Verbiest}, {Lentati}, {Hobbs}, {van
  Haasteren}, {Demorest}, {Janssen}, {Wang}, {Desvignes}, {Caballero}, {Keith},
  {Champion}, {Arzoumanian}, {Babak}, {Bassa}, {Bhat}, {Brazier}, {Brem},
  {Burgay}, {Burke-Spolaor}, {Chamberlin}, {Chatterjee}, {Christy}, {Cognard},
  {Cordes}, {Dai}, {Dolch}, {Ellis}, {Ferdman}, {Fonseca}, {Gair},
  {Garver-Daniels}, {Gentile}, {Gonzalez}, {Graikou}, {Guillemot}, {Hessels},
  {Jones}, {Karuppusamy}, {Kerr}, {Kramer}, {Lam}, {Lasky}, {Lassus},
  {Lazarus}, {Lazio}, {Lee}, {Levin}, {Liu}, {Lynch}, {Lyne}, {Mckee},
  {McLaughlin}, {McWilliams}, {Madison}, {Manchester}, {Mingarelli}, {Nice},
  {Os{\l}owski}, {Palliyaguru}, {Pennucci}, {Perera}, {Perrodin}, {Possenti},
  {Petiteau}, {Ransom}, {Reardon}, {Rosado}, {Sanidas}, {Sesana}, {Shaifullah},
  {Shannon}, {Siemens}, {Simon}, {Smits}, {Spiewak}, {Stairs}, {Stappers},
  {Stinebring}, {Stovall}, {Swiggum}, {Taylor}, {Theureau}, {Tiburzi},
  {Toomey}, {Vallisneri}, {van Straten}, {Vecchio}, {Wang}, {Wen}, {You},
  {Zhu}, \& {Zhu}}]{verbiest16}
{Verbiest}, J.~P.~W., {Lentati}, L., {Hobbs}, G., {et~al.} 2016,
  \href{http://dx.doi.org/10.1093/mnras/stw347}{\JournalTitle{\mnras}, 458,
  1267}

\bibitem[{{Vigeland} {et~al.}(2018){Vigeland}, {Islo}, {Taylor}, \&
  {Ellis}}]{vigeland18}
{Vigeland}, S.~J., {Islo}, K., {Taylor}, S.~R., \& {Ellis}, J.~A. 2018,
  \href{http://dx.doi.org/10.1103/PhysRevD.98.044003}{\JournalTitle{\prd}, 98,
  044003}

\bibitem[{{Wandelt} {et~al.}(2001){Wandelt}, {Hivon}, \&
  {G{\'o}rski}}]{wandelt01}
{Wandelt}, B.~D., {Hivon}, E., \& {G{\'o}rski}, K.~M. 2001,
  \href{http://dx.doi.org/10.1103/PhysRevD.64.083003}{\JournalTitle{\prd}, 64,
  083003}

\bibitem[{Waskom {et~al.}(2017)Waskom, Botvinnik, O'Kane, Hobson, Lukauskas,
  Gemperline, Augspurger, Halchenko, Cole, Warmenhoven, de~Ruiter, Pye, Hoyer,
  Vanderplas, Villalba, Kunter, Quintero, Bachant, Martin, Meyer, Miles, Ram,
  Yarkoni, Williams, Evans, Fitzgerald, Brian, Fonnesbeck, Lee, \&
  Qalieh}]{seaborn}
Waskom, M., Botvinnik, O., O'Kane, D., {et~al.} 2017,
  \href{http://dx.doi.org/10.5281/zenodo.883859}{\JournalTitle{mwaskom/seaborn:
  v0.8.1}}

\bibitem[{{Zhu} {et~al.}(2016){Zhu}, {Wen}, {Xiong}, {Xu}, {Wang}, {Mohanty},
  {Hobbs}, \& {Manchester}}]{zhu16}
{Zhu}, X.-J., {Wen}, L., {Xiong}, J., {et~al.} 2016,
  \href{http://dx.doi.org/10.1093/mnras/stw1446}{\JournalTitle{\mnras}, 461,
  1317}

\end{thebibliography}

\listofchanges

\end{document}